%% file: main.tex
\documentclass[manuscript]{acmart}

\AtBeginDocument{%
  \providecommand\BibTeX{{%
    \normalfont B\kern-0.5em{\scshape i\kern-0.25em b}\kern-0.8em\TeX}}}

\acmYear{2024}\copyrightyear{2024}
\setcopyright{acmlicensed}
\acmConference[ACM FAccT '24]{ACM Conference on Fairness, Accountability, and Transparency}{June 3--6, 2024}{Rio de Janeiro, Brazil}
\acmBooktitle{ACM Conference on Fairness, Accountability, and Transparency (ACM FAccT '24), June 3--6, 2024, Rio de Janeiro, Brazil}
\acmDOI{10.1145/3630106.3658547}
\acmISBN{979-8-4007-0450-5/24/06}

\usepackage{booktabs}
\usepackage{enumerate}
\usepackage{color}
\usepackage{comment}
\usepackage{enumitem}
\usepackage{url}
\usepackage{multirow}
\usepackage{MnSymbol}
\usepackage{hyperref}

\begin{document}

\fancyhead{}

\title{Why is ``Problems'' Predictive of Positive Sentiment? A Case Study of Explaining Unintuitive Features in Sentiment Classification}

\author{Jiaming Qu}
\affiliation{%
  \institution{University of North Carolina at Chapel Hill}
  \city{Chapel Hill}
  \state{North Carolina}
  \country{USA}}
\email{jiaming@unc.edu}

\author{Jaime Arguello}
\affiliation{%
  \institution{University of North Carolina at Chapel Hill}
  \city{Chapel Hill}
  \state{North Carolina}
  \country{USA}}
\email{jarguello@unc.edu}

\author{Yue Wang}
\affiliation{%
  \institution{University of North Carolina at Chapel Hill}
  \city{Chapel Hill}
  \state{North Carolina}
  \country{USA}}
\email{wangyue@unc.edu}

\input{tex/abstract.tex}



\maketitle

\input{tex/introduction}
\input{tex/related_work}
\input{tex/methods}
\input{tex/results}
\input{tex/discussion}
\input{tex/conclusion}

\input{tex/ethical_stmt}

\bibliographystyle{ACM-Reference-Format}
\bibliography{reference}

\newpage
\input{tex/appendix}

\end{document}

%% file: tex/abstract.tex
\begin{abstract}
Explainable AI (XAI) algorithms aim to help users understand how a machine learning model makes predictions.  To this end, many approaches explain which input features are most predictive of a target label.  However, such explanations can still be puzzling to users (e.g., in product reviews, the word ``problems'' is predictive of \emph{positive} sentiment).  If left unexplained, puzzling explanations can have negative impacts.  Explaining unintuitive associations between an input feature and a target label is an underexplored area in XAI research. We take an initial effort in this direction using unintuitive associations learned by sentiment classifiers as a case study. We propose approaches for (1) automatically detecting associations that can appear unintuitive to users and (2) generating explanations to help users understand why an unintuitive feature is predictive. Results from a crowdsourced study ($N=300$) found that our proposed approaches can effectively detect and explain predictive but unintuitive features in sentiment classification. 
\end{abstract}

%% file: tex/introduction.tex
\section{Introduction}

Research on explainable artificial intelligence (XAI) has investigated a variety of approaches to explaining the complex behavior of a machine learning model. One simple and straightforward approach is to show which parts of an input (i.e., which features) are most influential.  Such explanations are referred to as \emph{feature importance} explanations.  Different algorithms and visualizations have been developed to generate feature importance explanations~\cite{lime2016,sundararajan2017axiomatic,lundberg2017unified, bansal2021does,lai2019human}.  Such explanations have been empirically shown to improve a user's performance in a variety of tasks, such as AI-assisted decision-making~\cite{bansal2021does,lai2019human}.

Despite promising results, studies on feature importance explanations have not tackled an important problem---syntactically simple explanations (e.g., ``feature $x$ plays an important role in predicting category $y$'') can still be puzzling or counterintuitive.  For example, prior XAI research has found that having asthma lowers the risk of death among pneumonia patients~\cite{caruana2015intelligible}; that the word ``Chicago'' is a strong indicator of a Chicago hotel review being fake~\cite{lai2020chicago}; and that words like ``host'' and ``posting'' have a stronger association with Atheism than Christianity in a topical classification task~\cite{lime2016,lime2016github}. In these cases, features that are deemed ``important'' by a model may not immediately make sense to humans.  Most XAI approaches that focus on feature importance do not further explain \emph{why} a feature is important.  

We use the term \textbf{unintuitive features} to describe this phenomenon. Unintuitive features are predictive from a model's perspective but are at odds with human intuition and common sense.  An important question is: What makes a feature unintuitive?  There are at least two possibilities.  First, a predictive feature may be unintuitive because of anomalies in the training data, especially when training data is sparse and the feature is predictive due to overfitting. Second, a predictive feature may be unintuitive because it represents an \emph{underlying} phenomenon that is not obvious to a human by simply seeing an explanation such as ``feature $x$ is predictive of category $y$''.  For example, within the context of automotive product reviews, the word ``fit'' (a seemingly positive word) is predictive of negative sentiment. At first, this seems paradoxical. It even seems that the classifier learned an incorrect association. However, this is not the case. The word ``fit'' predicts negative sentiment because people tend to use it when the product did not ``fit''. Conversely, people do not use ``fit'' in positive reviews because a product ``fitting'' is a minimum requirement unworthy of mentioning in a positive review.  In our research, we focus on the second category and not the first.  That is, we focus on features that are: (1) predictive from a model's perspective, (2) generalizable to test data, and (3) unintuitive to a human.

Prior work has mostly focused on algorithms that can translate a complex model's predictive behavior into syntactically simple forms~\cite{lime2016,ribeiro2018anchors,sundararajan2017axiomatic,lundberg2017unified,shrikumar2017learning}, which is a necessary but not sufficient condition for algorithm-generated explanations to make sense to humans.  Most studies that have evaluated feature importance explanations have assumed that syntactically simple explanations are self-explanatory. Lai et al.~\cite{lai2020chicago} touched upon this issue to some extent (e.g., using manually curated rules to further explain why words like ``Chicago'' are predictive of a Chicago hotel review being fake).  However, they did not provide an algorithmic solution for generating further explanations.

Explaining unintuitive features is an important problem in XAI research. Prior studies have found that unintuitive explanations can make users lose trust in a machine learning model~\cite{cai2019human,oh2020understanding,qu2023,chen2023understanding}.  Additionally, if left unexplained, users may hypothesize wrong reasons for why an unintuitive feature is predictive.  Schuff et al.~\cite{schuff2022human} randomly highlighted words within reviews as being predictive of a sentiment.  Results found that participants made up their own incorrect explanations for why those words were predictive.  Thus, explaining unintuitive features may improve users' trust in a model and help them learn about the predictive task.

In this paper, we take initial steps toward addressing the issue of unintuitive features in XAI.  As a case study, we focus on a sentiment classification task---predicting whether an Amazon product review is positive or negative. It is a task that can be performed by ordinary crowdworkers and machine learning models can perform reasonably well on, and therefore unintuitive features are not due to overfitting. We identified words that are predictive of a specific sentiment but likely to be perceived as unintuitive or paradoxical to a human.  For example, the word ``problems'' (a seemingly negative word) is predictive of \emph{positive} sentiment and the word ``fit'' (a seemingly positive word) is predictive of \emph{negative} sentiment. We report on a crowdsourced user study ($N=300$) that evaluated different tools designed to explain the predictiveness of an unintuitive feature.  Participants were assigned to one of six interface conditions (a between-subjects design). Interface conditions varied based on the tools available to participants.    The study investigated three research questions, which considered the effects of the interface condition on different types of dependent variables:

\begin{itemize}[leftmargin=*]
\item \textbf{RQ1:}  How does the interface condition affect participants' \emph{understanding} of an unintuitive feature's predictiveness? 
\item \textbf{RQ2:}  How does the interface condition affect participants' \emph{perceptions} of the provided tools and their experiences?
\item \textbf{RQ3:}  How does the interface condition affect participants' \emph{behaviors} during different tasks?
\end{itemize}

The study proceeded in two phases.  \textsc{Phase 1} of the study validated our assumption that the level of (un)intuitiveness of a predictive feature in sentiment classification can be computationally estimated. During \textsc{Phase 1},  participants were shown a batch of predictive features (i.e., words) and asked to judge which sentiment they expected the word to convey: positive, negative, or ``not sure''. These judgments were found to strongly correlate with those made by a large language model. \textsc{Phase 2} of the study investigated the above three research questions. During \textsc{Phase 2}, participants completed four trials. During each trial, participants were shown a predictive but unintuitive word and asked to complete different judgments and tasks using the tools available in their assigned interface condition.  We explored three different tools: (1) a visualization of the sentiment label distribution among training instances containing the word, (2) training examples of either sentiment containing the word, and (3) contextual patterns minded from training examples of either sentiment containing the word.

Our results found that participants had the best outcomes when provided with a combination of tools (data distribution + examples or contextual patterns).  When provided with \emph{only} the data distribution tool, participants were able to correctly judge the sentiment of the unintuitive features.  However, they did not perceive the tool to be understandable, helpful, nor trustworthy.  Seeing concrete examples and contextual patterns helped participants explain \emph{why} an unintuitive feature is predictive. Through the case study, our paper contributes practical tools and design implications for explaining predictive yet unintuitive features, an important but underexplored area in XAI research.

%% file: tex/related_work.tex
\section{Related Work}

Our research builds upon three areas of prior work: (1) technical approaches for explaining a model's predictions, particularly feature importance approaches, (2) prior studies where machine-generated explanations were found to be unintuitive or puzzling, and (3) empirical studies that evaluated XAI systems through user experiments.

\textbf{Feature Importance Explanations:} Explainable artificial intelligence (XAI) research has explored a wide range of approaches to help people understand a machine learning model's predictions. For example, to explain a prediction for a specific instance, approaches can highlight which features of the instance are the most indicative of the predicted label~\cite{lime2016,ribeiro2018anchors,sundararajan2017axiomatic,lundberg2017unified}, which training instances are the most influential in teaching the model to predict the label for this instance~\cite{koh2017understanding,yeh2018representer,pruthi2020estimating}, and which training instances are most similar to the target instance and have the same ground truth label as the prediction~\cite{cai2019effects,yang2020visual}. Among these approaches, feature importance (or feature attribution) explanations are highly popular. Such explanations highlight which parts of the input (e.g., words, sentences, superpixels) are most indicative of the predicted label~\cite{lime2016,sundararajan2017axiomatic,lundberg2017unified,ribeiro2018anchors}.

In XAI research, explanations can be categorized as either global or local.  Global explanations provide insights about the overall behavior of the model, while local explanations elucidate a model's predictions on individual instances~\cite{doshi2017towards}.  This distinction also applies to feature importance explanations. Global feature importance explanations show a feature's overall impact in the model's prediction logic, which is typically learned from the entire training data. These explanations highlight which features have the most influence on the model's predictions across a wide range of inputs~\cite{covert2020understanding}. Some machine learning models have mechanisms that can be leveraged for showing global feature importance, such as the coefficients from a support vector machine model~\cite{lai2019human,lai2020chicago} or logistic regression model~\cite{bursac2008purposeful,carton2020feature}. In our study, we trained logistic regression classifiers and used regression coefficients to identify predictive (i.e., important) features.
Compared to global explanations, local feature importance explanations are more intricate because they explain the importance of features for a specific instance. These explanations demonstrate why a model made a certain prediction for a given input. Prior research has developed different algorithms to compute feature importance locally, e.g., LIME~\cite{lime2016}, Integrated Gradients~\cite{sundararajan2017axiomatic}, and SHAP~\cite{lundberg2017unified}. Despite their differences, global and local feature importance methods share the same syntax: ``feature $x$ plays an important role in predicting label $y$ for a specific instance or any label $y \in Y$ across instances'' and assume that users can understand a model's prediction based on certain important features. However, studies have found that this assumption is not always true.  That is, users are sometimes confused after learning which features play an important role. We discuss such studies below.

\textbf{The Phenomenon of Unintuitive Explanations:} Machine-generated explanations are intended to help users understand a model's behavior.  However, such explanations are not guaranteed to always make sense to users.  The phenomenon of unintuitive explanations is not rare in previous empirical studies. For example, Qu et al.~\cite{qu2023} conducted a study in which participants scrutinized a document's machine-predicted categories by inspecting the most influential sentences, highlighted by the system.  Participants commented on ignoring such explanations when they could not understand why the sentences were important. Unclear explanations can also have undesirable consequences for domain experts.  In a study where pathologists completed a diagnostic task assisted by example-based explanations, participants exhibited confusion and self-doubt when they did not understand or agree with the explanations~\cite{cai2019effects}. 

Prior studies have also observed unintuitive feature importance explanations.  For example, studies have reported unexpected regression coefficients in analyses related to econometrics~\cite{farrar1967multicollinearity,kennedy2005oh}, psychology~\cite{kaufmann1986interactions}, and education~\cite{nathans2012interpreting}. 
More recently, XAI research has reported predictive but unintuitive features in the medical domain (e.g., patients with pneumonia who have a history of asthma have a lower risk of death)~\cite{caruana2015intelligible} and text analysis domain (e.g., the word ``Chicago'' is a strong predictor of a review being fake)~\cite{lai2020chicago}.
One solution to addressing unintuitive features is to consider feature interactions instead of single features~\cite{tsang2018can,borisov2022relational,janizek2021explaining}. However, feature interaction explanations can only resolve simple unintuitive cases such as negation because they focus on interactions between \emph{pairs} of features~\cite{borisov2022relational,janizek2021explaining}.  Text analysis often involves predicting complex phenomena (e.g., topic, sentiment, and intent).  Such phenomena are abstract and can be influenced by patterns that go beyond pairwise interactions between words. This inspires us to develop tools to explain unintuitive text features in more complex situations by mining semantic patterns from the training data.

\textbf{Empirical Studies in XAI:} Evaluating XAI systems with human end-users gives direct evidence on the effectiveness of explanations in a concrete task scenario~\cite{doshi2017towards,zhou2021evaluating}. To this end, numerous empirical studies have been conducted to investigate human-XAI interaction using a variety of tasks across different domains such as sentiment analysis~\cite{bansal2021does}, topic categorization~\cite{qu2023}, deceptive review detection~\cite{lai2020chicago}, disease diagnosis~\cite{cai2019human}, and toxicity detection~\cite{carton2020feature}. To gauge the effects of XAI systems on human end-users, previous research conducted both quantitative and qualitative evaluations. Quantitative evaluations often measured end-users' (1) performance (e.g., decision accuracy~\cite{bansal2021does,qu2023,lai2020chicago,cai2019human,carton2020feature}), (2) perceptions (e.g., confidence~\cite{chromik2021think,cai2019human}, understanding~\cite{cai2019effects,qu2023} and trust~\cite{cai2019human,lai2020chicago}), and (3) behaviors (e.g., time spent on task~\cite{qu2023,cai2019human,carton2020feature}). Besides quantitative evaluations, prior studies have also used qualitative techniques to gain deeper insights into human-XAI interaction, such as conducting exit interviews with participants~\cite{qu2023,chromik2021think,cai2019human} or using a think-aloud protocol~\cite{oh2020understanding,buccinca2020proxy,chen2023understanding}. In our study, we collected both quantitative and qualitative data to investigate whether our developed tools helped participants understand the predictiveness of an unintuitive feature.

%% file: tex/methods.tex
\section{Methods}

\subsection{Study Overview}
To investigate RQ1-RQ3, we conducted a crowdsourced study using the Prolific platform.
The study involved 300 participants ($M=116$, $F=183$, $Unreported=1$). Participants' ages ranged from  18 to 63 ($Mean = 35.41$, $S.D. = 11.92$).  We restricted our study to English-speaking Prolific workers from USA, UK, and Canada who had completed at least 100 tasks with an acceptance rate $\ge$ 95\% and had experience in online shopping and review writing. The study involved two phases: \textsc{Phase 1} and \textsc{Phase 2}. 

During \textsc{Phase 1}, participants were shown a list of 10 words.  For each word, participants were asked to indicate which sentiment they expected the word to convey: ``positive'', ``negative'' or ``not sure''.  \textsc{Phase 1} used the same interface for all participants.  Our goal for \textsc{Phase 1} was to investigate whether an LLM-based zero-shot classifier can automatically estimate the (un)intuitiveness of a word that is predictive of positive or negative sentiment in product reviews. Each word was redundantly classified by five participants.  This enabled us to compare the level of disagreement among participants with the level of (un)intuitiveness estimated using the LLM-based zero-shot classifier.

During \textsc{Phase 2}, participants completed four trials. During each trial, participants were shown a predictive yet unintuitive feature and were asked to complete several tasks and answer several questions (e.g., judge whether the feature is predictive of positive or negative sentiment).  While \textsc{Phase 1} used the same interface for all participants, \textsc{Phase 2} involved an interface manipulation.  Participants were assigned to 1 of 6 interface conditions (50 participants per condition).  Interface conditions varied based on the combination of different tools that we designed to explain the predictiveness of an unintuitive feature (Section~\ref{subsec:conditions}).  After completing all four trials, participants completed a questionnaire that asked about their perceptions of the system and their experiences (Section~\ref{subsec:rq2_perceptions}). Participants were given US\$ 6.00 for participation.  The study was approved by our Institutional Review Board (IRB).

\subsection{Identifying and Explaining Unintuitive Features in Sentiment Classification}
\label{subsec:data_algos}

\textbf{Dataset and Models:} The data used in our study originated from Ni et al.~\cite{ni2019justifying}, which consists of Amazon product reviews of various product categories. We selected five product categories: Automotive, Electronics, Pet Supplies, Home and Kitchen, and Sports and Outdoors. 
In the original dataset, each review had a 5-star rating. We used reviews with 1 star as negative and 5 stars as positive. For each product category, we trained a logistic regression classifier using a balanced dataset of 200,000 reviews. All classifiers used a unigram TF-IDF representation with stopwords removed~\cite{bird2009natural,pedregosa2011scikit}. Each classifier was tested on a balanced dataset of 10,000 reviews from the same product category.  All classifiers achieved an F1 score $\ge$ 0.90. 

\textbf{Identifying Predictive Features:} Unigram feature importance explanations are widely used in prior XAI research~\cite{kulesza2015principles, nguyen2018comparing, lai2020chicago}. In this study, we identified predictive features in the above logistic regression classifiers. For each product category, we selected the 200 words with the highest coefficients as the most predictive of positive sentiment (denoted as $\mathcal{S}^+$) and the 200 words with the lowest coefficients as the most predictive of negative sentiment (denoted as $\mathcal{S}^-$).  Words shown to participants during \textsc{Phase 1} and \textsc{Phase 2} were sampled from these sets. 

\textbf{Estimating the (Un)intuitiveness of a Predictive Feature:}  Predictive features can have different levels of (un)intuitiveness.  For example, it is obvious why ``great'' is predictive of positive sentiment and ``terrible'' is predictive of negative sentiment.  However, it is unclear why ``problems'' (a seemingly negative word) is predictive of \emph{positive} sentiment and ``fit'' (a seemingly positive word) is predictive of \emph{negative} sentiment.  One approach to estimating the (un)intuitiveness of a word-sentiment relation would be through human assessment.  However, this requires significant manual effort.  
Instead, we leveraged a large language model (LLM) to estimate whether a word-sentiment relation might be perceived as (un)intuitive to humans.  

The basic idea is to use an LLM to approximate a human's perception that a word $w$ conveys a sentiment $y$. We used an LLM called BART~\cite{lewis2020bart}. BART can predict the probability that one piece of text logically entails another. In this respect, it can be used as a classifier when one piece of text is the input and the other is the textual description of a candidate label (e.g., positive or negative sentiment). In this setup, the LLM is used as a ``zero-shot classifier’’ because it does not require training data~\cite{yin2019benchmarking}. To estimate the (un)intuitiveness of a word-sentiment relation, we used the prompt ``\texttt{In Amazon reviews of [CATEGORY] products, word $w$ is $y$}'' and asked the zero-shot classifier to predict the probabilities of class labels $y=$ [``positive'', ``negative''] for word $w$.\footnote{We used the implementation in the HuggingFace library: https://huggingface.co/tasks/zero-shot-classification.} We use $P_z(y|w)$ to denote the probability that word $w$ conveys sentiment $y$ according to the zero-shot classifier. A large value of $P_z(y=pos|w)$ suggests that $w$ is intuitively predictive of positive sentiment.  A large value of $P_z(y=neg|w) = 1 - P_z(y=pos|w)$ suggests that $w$ is intuitively predictive of negative sentiment.  Values close to 0.5 suggest that $w$ is not intuitively associated with either positive or negative sentiment. By adopting this approach, we assume that the zero-shot classifier approximates a human’s perception that word $w$ conveys sentiment $y$. \textsc{Phase 1} of our user study validated this assumption (Section~\ref{results_validating_llm}).

\textbf{Explaining Unintuitive Features:} We developed three different tools to explain the relation between a word $w$ and a sentiment $y$. Our tools explained a word's association with \emph{both} sentiment labels (i.e., positive and negative).  During \textsc{Phase 2}, participants were asked to scrutinize both associations and choose the one that made the most sense to them.

The first tool, \textbf{\textsc{Distribution}},  was designed to show posterior probabilities $P_D(y=pos|w)$ and $P_D(y=neg|w)$, inspired by prior work~\cite{kulesza2015principles}.  These probabilities were estimated based on the proportion of positive and negative reviews containing word $w$ in the training data $D$. We used a pie chart to visualize these probabilities. 

The second tool, \textbf{\textsc{Example}}, was designed to show training examples where $w$ and $y$ co-occur, another common approach in prior work~\cite{lai2020chicago}. We sampled 25 positive and 25 negative reviews containing $w$ from the training data. 

The third tool, \textbf{\textsc{Pattern}}, was designed to show \emph{contextual patterns} where $w$ and $y$ co-occur. Contextual patterns are common phrases of variable lengths that appear in the training data. For example, the word ``minutes’’ has positive contextual patterns such as ``5 minutes to install'' and negative contextual patterns such as ``broke within minutes''.  To show contextual patterns for word $w$ and sentiment $y$, we developed a \emph{contextual pattern mining} algorithm, a novel contribution of this work.  Given word $w$ and sentiment $y$, the algorithm identifies common and diverse patterns (i.e., phrases) $p$ in the training data that satisfy two conditions: (1) pattern $p$ contains word $w$ and (2) $P_z(y|p)$ is large (i.e., phase $p$ clearly predicts sentiment $y$ according to the zero-shot classifier). We describe the algorithm below.

Given word $w$ and sentiment $y$, the first step is to find candidate contextual patterns that include $w$ and are predictive of $y$.  To this end, we first iterate over all training instances associated with sentiment $y$ that contain word $w$.  For each instance, we consider phrases of increasing length by adding words to the left and right of $w$.  We limit ourselves to phrases no longer than five words to the left and right of $w$.  For each newly-generated phrase, we use the zero-shot classifier to estimate $P_z(y|p)$, the probability that the phrase $p$ is predictive of sentiment $y$.  The shortest phrase $p$ that gives $P_z(y|p) > 0.8$ (if any) is then considered a \emph{candidate} contextual pattern for $w$ and $y$.
The next step is to select a small set of candidate contextual patterns to show for $w$ and $y$.  Inspired by the maximal marginal relevance (MMR) algorithm~\cite{carbonell1998use}, we select the most frequent pattern first.  Then, we iteratively select patterns that are both frequent and semantically dissimilar to previously selected patterns. To measure semantic similarity, we used cosine similarity between phrase embeddings computed by a transformer-based encoder~\cite{reimers2019sentence}. This selection process ensured that participants were exposed to contextual patterns that are both predictive of $y$ and diverse. Table~\ref{table:unintuitive_words_patterns_examples} shows example positive and negative contextual patterns mined for words that were estimated as unintuitive by the zero-shot classifier.

\begin{table*}[]
\caption{Examples of contextual patterns mined for unintuitive words. Words like ``problems'' and ``minutes'' are estimated to be unintuitively positive, while words like ``star'' and ``money'' are estimated to be unintuitively negative. }
\begin{tabular}{ccc}
\toprule
Word     & {{\color{orange}Positive}} contextual patterns & {{\color{blue}Negative}} contextual patterns \\ \midrule
problems &  no more problems, without any problems & problems with, problems since   \\
minutes  &  5 minutes to install, installs in minutes &  broke within minutes, didn't last 5 minutes  \\ \midrule
star     &  star rating, 5 star, star product  &   one star, half star, negative star \\ 
money    &  worth the money, good value for money & waste of money, want my money back  \\ \bottomrule
\end{tabular}
\label{table:unintuitive_words_patterns_examples}
\end{table*}

\subsection{Experimental Design}\label{subsec:exp_design}

In this section, we describe how words were sampled for \textsc{Phase 1} and \textsc{Phase 2}. Our goal of \textsc{Phase 1} was to validate the use of a zero-shot classifier to estimate the (un)intuitiveness of a feature that is predictive of positive or negative sentiment.  During \textsc{Phase 1}, we wanted to expose participants to predictive words with different levels of (un)intuitiveness.  For each product category, we used the zero-shot classifier to estimate $P_z(y=pos|w)$ for every word in sets $\mathcal{S}^+$ and $\mathcal{S}^-$ (i.e., predictive of positive and negative according to the logistic regression model).  We only estimated $P_z(y=pos|w)$ since $P_z(y=neg|w) = 1 - P_z(y=pos|w)$.  Then, we sampled 120 words associated with different levels of (un)intuitiveness using stratified sampling. These 120 words were organized into 12 batches of 10 words each.  Each batch was redundantly judged by five participants.  These redundant judgements were used to estimate $P_u(y|w)$, the probability that word $w$ predicts sentiment $y$ according to human users.  As described in Section~\ref{results_validating_llm}, a significant correlation between $P_u(y|w)$ and $P_z(y|w)$ validated our use of a zero-shot classifier to estimate the (un)intuitiveness of a predictive feature.

During \textsc{Phase 2}, our goal was to investigate how different tools can help people understand the predictiveness of an unintuitive feature.  For each product category, we sampled 40 words from sets $\mathcal{S}^+$ and $\mathcal{S}^-$ as follows.  First, we excluded words sampled for \textsc{Phase 1}.  Then, we sampled words that met the following criteria.  First, we included words with $P_z(y=pos|w) < 0.2$ from set $\mathcal{S}^+$.  These are words that are \emph{paradoxically} predictive of positive (e.g., the word ``problems'' being predictive of positive).  Second, we included words with $P_z(y=pos|w) > 0.8$ from set $\mathcal{S}^-$.  These are words that are \emph{paradoxically} predictive of negative (e.g., the word ``fit'' being predictive of negative).  Third, we included words with $0.2 \le P_z(y=pos|w) \le 0.8$ from sets $\mathcal{S}^+$ and $\mathcal{S}^-$.  These are words that are unintuitive regardless of which sentiment they are predictive of.  Finally, during \textsc{Phase 2}, participants were asked whether an AI system should consider the word as strong predictive evidence of sentiment.  Therefore, we only included words that resulted in a statistically significant drop in performance if omitted from a logistic regression model.  For each product category, each sample of 40 words was organized into 10 batches of 4 words each (2 from $\mathcal{S}^+$ and 2 from $\mathcal{S}^-$). Each batch was completed by six participants, each in a different interface condition. Sixty participants were assigned to each product category. More details of the study task allocation are available in Appendix~\ref{appendix_a}.

\subsection{Study Protocol}\label{subsec:protocol}

The study protocol proceeded as follows.  First, participants watched an overview video of the study.  Then, participants completed \textsc{Phase 1} of the study as follows.  After watching an instructional video, participants were presented with a batch of 10 words and were instructed to judge the sentiment conveyed by each word.  Participants were presented with the following prompt: ``Think about Amazon reviews for [Category] products. Which sentiment is each word more likely to convey?''  Participants were instructed to select the sentiment of each word based solely on their intuition.  Participants were presented with the options of ``positive, ``negative'', and ``not sure''.

Next, participants proceeded to \textsc{Phase 2} of the study.  During \textsc{Phase 2}, participants completed four trials.  During each trial, participants were exposed to an unintuitive feature and were asked to complete a series of judgments.  While \textsc{Phase 1} used the same interface for all participants, \textsc{Phase 2} involved an interface manipulation.  Participants were assigned to one of six interface conditions (a between-subjects design).  Interface conditions varied based on the tools available to participants. For each word, participants were asked to complete a series of judgments. First, participants were asked to judge the sentiment of the word.  Participants used a range slider to indicate their perceived sentiment from very negative to very positive. The range slider did not have a midpoint. Therefore, participants had to choose between positive or negative.  However, they could choose values close to the midpoint if they were unsure. Second, participants were prompted to list scenarios in which the word might be used to convey the selected sentiment.  Participants were instructed that ``scenarios can be phrases, sentences, or explanations based on your personal understanding.'' Participants were provided with a textbox to list scenarios as a bulleted list. Third, participants were asked to rate their confidence in their responses to the first two tasks on a 5-point scale.  Next, participants were asked whether the AI system should consider the word as \emph{strong} evidence of the sentiment they selected.  Participants were asked to respond ``yes'' or ``no'' using radio buttons.  Given that all words sampled for \textsc{Phase 2} were highly predictive, resulting in statistically significant drops in performance if omitted, the correct answer was always ``yes''.  However, participants did not know this. Then, participants were asked to rate their confidence in their response to the above question on a 5-point scale. After all four \textsc{Phase 2} trials, participants were presented with a side-by-side comparison between the AI system's judgment of each word (based on regression coefficients) and their own judgment. Finally, participants completed a post-task questionnaire about their perceptions of the interface and the task. Our study materials and system demos are \href{https://jiamingqu.com/FAccT24_demo/}{\textcolor{blue}{\underline{available online}}}.

\subsection{Phase 2 Interface Conditions}\label{subsec:conditions}
\begin{figure*}[htbp]
\centering
\includegraphics[width=\textwidth]{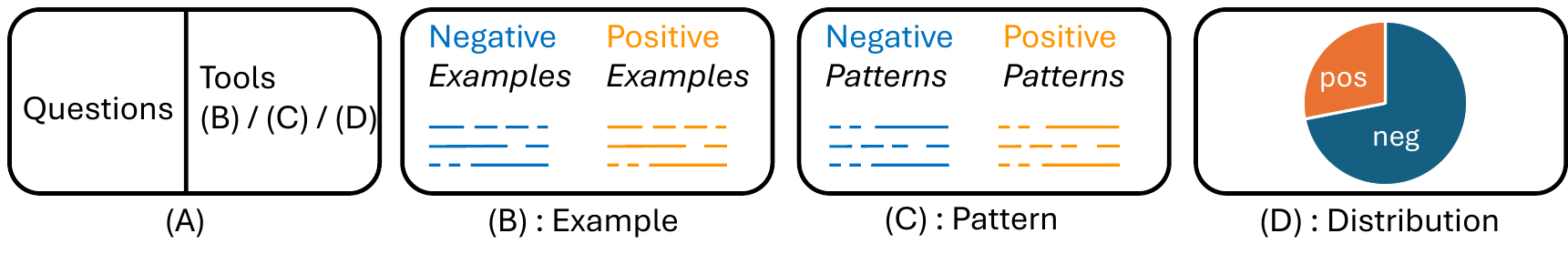}
\caption{\textsc{Phase 2} interface design.  Questions and tools (if any) were displayed side-by-side (A). Visual representations of our tools are shown in subfigures B-D.
}
\label{fig:phase2_screenshot}
\end{figure*}

In \textsc{Phase 2}, participants were randomly assigned to one of six interface conditions (i.e., a between-subjects design). Interface conditions varied based on the tools available to participants.  Participants answered the same questions in all interface conditions. Figure~\ref{fig:phase2_screenshot} describes the layout of the interface (A) and our three tools (B-D). More details of the interface are available in Appendix~\ref{appendix_b}.

\textbf{\textsc{Baseline}:} In this condition, no tools were provided. Participants made judgments based solely on their intuition.

\textbf{\textsc{Example} (Figure~\ref{fig:phase2_screenshot}-B):} In this condition, we randomly sampled 25 positive and 25 negative reviews from the training data containing the word to be judged.

\textbf{\textsc{Pattern} (Figure~\ref{fig:phase2_screenshot}-C):} In this condition, we used the \emph{contextual pattern mining} algorithm (Section~\ref{subsec:data_algos}) to display positive and negative contextual patterns associated with the word to be judged. We displayed up to five contextual patterns per sentiment and provided three sampled reviews per pattern-sentiment pair.

\textbf{\textsc{Distribution} (Figure~\ref{fig:phase2_screenshot}-D):} In this condition, we provided the sentiment label distribution of training instances containing the word to be judged.

\textbf{\textsc{Example+Distribution}:} In this condition, participants had access to both lists of reviews and label distribution (i.e., both (B) and (D) in Figure~\ref{fig:phase2_screenshot}).  Compared to the \textsc{Example} condition, this condition also displayed the number of positive and negative reviews containing the word. Lists of reviews were shown by default, and participants could use radio buttons to switch between tools.

\textbf{\textsc{Pattern+Distribution}:} In this condition, participants had access to both contextual patterns and label distribution (i.e., both (C) and (D) in Figure~\ref{fig:phase2_screenshot}). Compared to the \textsc{Pattern} condition, this condition also displayed the number of reviews associated with each contextual pattern. Contextual patterns were shown by default, and participants could use radio buttons to switch between tools.

\subsection{Measures of Understanding (RQ1)}\label{subsec:rq1_understand}
In RQ1, we investigated the effects of the interface condition on participants' understanding of a predictive but unintuitive feature. We measured participants' understanding from three perspectives.

\textbf{Correctness of Sentiment Judgment:} This binary measure considered whether a participant's judgment of a word's sentiment (i.e., positive vs.~negative) aligned with the logistic regression classifier.

\textbf{Correctness of Feature Consideration:} This binary measure considered: (1) whether the participant's judgment of the word's sentiment (i.e., positive vs.~negative) aligned with the logistic regression classifier and (2) whether the participant correctly indicated that the AI system should consider the word as strong evidence. Based on an ablation analysis, the correct answer for words sampled for \textsc{Phase 2} was always ``yes'' (i.e., the AI system should consider the word as strong evidence).  However, participants did not know this.

\textbf{Correctness of Listed Scenarios:} This measure considered the \emph{proportion} of valid scenarios that participants listed to support their selected sentiment. To this end, we conducted a qualitative analysis of scenarios listed by participants.  Each scenario was classified as ``valid'' if it met the following three criteria: (1) the scenario is relevant to the context of product reviews; (2) the scenario includes the word being judged; and (3) the scenario is relevant to the sentiment selected by the participant.  Our qualitative analysis of scenarios involved developing a coding guide.  After developing an initial coding guide, three of the authors annotated 30 lists of scenarios.  Then, the authors discussed disagreements and refined the coding guide.  Finally, to validate the coding guide, the same three authors annotated 30 new lists of scenarios.  The Fleiss' Kappa agreement was $\kappa = 0.5816$, which is considered moderate agreement~\cite{landis1977agreement}.  Finally, one author coded all remaining scenarios.  In total, participants listed 3,447 scenarios.

\subsection{Measures of Perceptions (RQ2)}\label{subsec:rq2_perceptions}
In RQ2, we investigated the effects of the interface condition on participants' perceptions of the interface and their experiences. Our first two measures are referred to as \textbf{Confidence in Sentiment Judgment} and \textbf{Confidence in Feature Consideration}. The first measure corresponds to the participant's confidence in judging the sentiment of a word and listing scenarios in which the word might be used to convey the selected sentiment.  The second measure corresponds to the participant's confidence in deciding whether the AI system should consider the word as strong predictive evidence. In both cases, participants rated their confidence on a 5-point scale ranging from (1) ``not at all confident'' to (5) ``extremely confident''.

After all four trials in \textsc{Phase 2}, participants completed a post-task questionnaire. Participants responded to agreement statements on a 7-point scale ranging from (1) ``strongly disagree'' to (7) ``strong agree''. The first part of the questionnaire asked four questions about: (1) \textbf{agreement} with the AI system's judgement of all four words, (2) \textbf{understanding} of the AI system's judgements and explanations, (3) \textbf{helpfulness} of the tools provided, and (4) \textbf{trust} in the AI system's effectiveness in predicting sentiment. The second part of the questionnaire asked about \textbf{system usability}.  We used the 10-item System Usability Scale (SUS)~\cite{brooke2013sus}. Responses to all 10 items had high internal consistency (Cronbach's $\alpha = 0.91$).  Therefore, we averaged responses to these 10 items to form one system usability measure. The third part of the questionnaire asked about \textbf{workload}.  We used the 6-item NASA-TLX~\cite{hart1988development}. Responses to these items had low internal consistency (Cronbach's $\alpha = 0.45$).  Therefore, we analyzed responses to these 6 items individually.

\subsection{Measures of Behaviors (RQ3)}\label{subsec:rq3_behaviors}
In RQ3, we investigated the effects of the interface condition on participants' behaviors. We considered three measures.

\textbf{Intensity in Sentiment Judgment:} This binary measure considered whether a participant made an extreme positive or negative judgment by choosing either the rightmost or leftmost positions on the range slider.

\textbf{Time Interval (sentiment judgment):} This measure considered the amount of time (in seconds) participants took to judge the sentiment of a word.

\textbf{Time Interval (all questions):} This measure considered the total amount of time (in seconds) participants took to answer all questions related to a word.

\subsection{Statistical Analysis}\label{subsec:statistics}

RQ1-RQ3 considered the effects of the interface condition on different types of outcomes.  To test for statistically significant effects, we fit linear regression models for real-valued measures and logistic regression models for binary measures.  Additionally, some measures involved four values originating from the four \textsc{Phase 2} trials. For such measures, we used multi-level modeling and added the participant ID as a random factor. In all models, we compared interface conditions against the \textsc{Baseline} condition to study the effects of providing \emph{any} tools for explaining unintuitive features versus providing \emph{none}. Non-\textsc{Baseline} conditions were included in each model as indicator variables. 

%% file: tex/results.tex
\section{Results}

\subsection{Validating the Use of an LLM-based Zero-shot Classifier to Estimate Feature (Un)intuitiveness }\label{results_validating_llm}

In our study, we leveraged an LLM-based zero-shot classifier to estimate whether a predictive feature is perceived as (un)intuitive to a human.  This approach assumes that the LLM can approximate a human's intuition about the relation between a word and a sentiment. To validate this assumption, we compared $P_z(y=pos|w)$ and $P_u(y=pos|w)$ across all 600 unique words judged during \textsc{Phase 1}.  $P_z(y=pos|w)$ denotes the probability that word $w$ predicts positive according to the zero-shot classifier. $P_u(y=pos|w)$ denotes the probability that word $w$ predicts positive according to our participants.  During \textsc{Phase 1}, each word $w$ was judged by five redundant participants using the options of ``positive'', ``negative'', or ``not sure''.  We estimated $P_u(y=pos|w)$ using the formula: $P_u(y=pos|w) = \frac{1 \cdot n_{pos} + 0.5 \cdot n_{ns} + 0 \cdot n_{neg}}{5}$.  We use $n_{pos}$, $n_{neg}$, and $n_{ns}$ to denote the number of participants who selected ``positive'', ``negative'' and ``not sure'' for word $w$, respectively. Essentially, this formula does a weighted aggregation of participants' judgments.  Then, we computed the Pearson correlation coefficient ($\rho$) between $P_z(y=pos|w)$ and $P_u(y=pos|w)$. The result showed a significant correlation ($\rho = 0.9125, p < .001$).  This high and significant correlation suggests that an LLM-based zero-shot classifier can estimate the extent to which a word-sentiment relation will be perceived as (un)intuitive to a human. More broadly, it suggests the possibility of using an LLM as a ``surrogate average user'' to automatically detect situations where XAI outputs are not self-explanatory and  further explanations are warranted.

\subsection{RQ1: Understanding}\label{results_understanding}

\begin{figure*}[htbp]
\centering
\includegraphics[width=\textwidth]{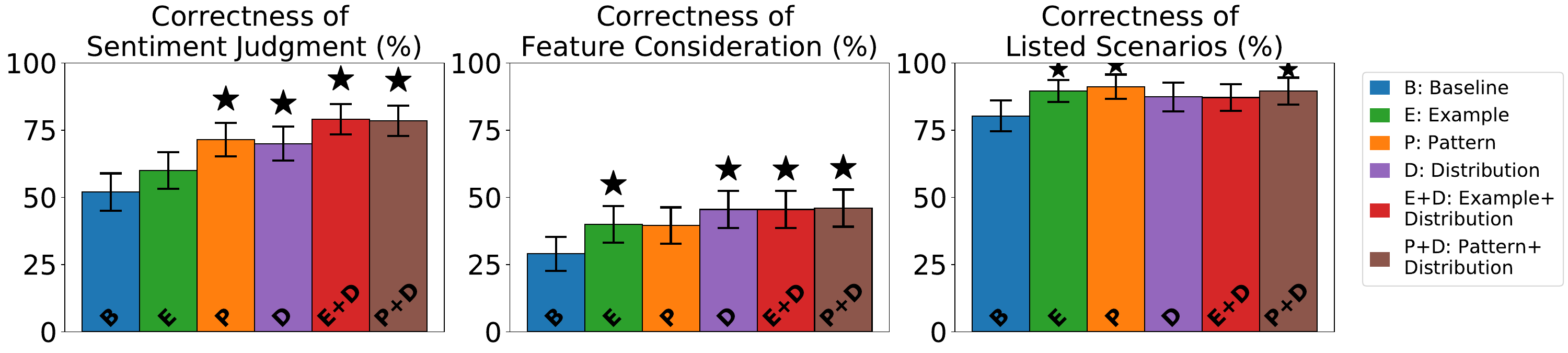}
\caption{Effects of different interface conditions on participants' \textbf{understanding} with means and 95\% confidence intervals. The star mark highlights interface conditions with statistically significant effects ($p < .05$) compared to the \textbf{\textsc{Baseline}} condition.}
\label{fig:understanding}
\end{figure*}

In RQ1, we investigated the effects of different interface conditions on participants' understanding of the predictiveness of an unintuitive feature. Figure~\ref{fig:understanding} shows our RQ1 results. Our results found three main trends.

First, in the \textsc{Baseline} condition, participants achieved 50\% accuracy in terms of ``correctness of sentiment judgement'' and 25\% accuracy in terms of ``correctness of feature consideration''.  In both cases, performance was not better than random guessing. Without our tools, only half of participants judged a word's sentiment correctly. Of these, only half correctly indicated that the model should consider the feature as strong predictive evidence. This result confirms that features selected for \textsc{Phase 2} indeed appeared unintuitive to participants. As a result, in the \textsc{Baseline} condition, their judgments approximated random guessing.

Second, all interface conditions, except the \textsc{Example} condition, had significant effects on participants making more correct sentiment judgments. All interface conditions, except the \textsc{Pattern} condition, had significant effects on participants making more correct feature consideration judgments. However, when participants had access to the label distribution in the other three interface conditions, they made significantly more correct judgments in both questions. Compared to providing a single tool, providing a combination of tools (i.e., the \textsc{Example+Distribution} and \textsc{Pattern+Distribution} conditions) was the best approach.

Third, compared to the \textsc{Baseline} condition, participants having access to concrete explanations in the \textsc{Example} and \textsc{Pattern} conditions were significantly more likely to list valid scenarios. Conversely, participants had difficulty listing valid scenarios in the \textsc{Distribution} condition, where they were only shown the label distribution. Moreover, showing grouped examples had greater effects than ungrouped examples---the \textsc{Pattern+Distribution} condition significantly increased the chance of participants listing valid scenarios but the \textsc{Example+Distribution} condition did not.

\subsection{RQ2: Perceptions}

\begin{figure*}[htbp]
\centering
\includegraphics[width=\textwidth]{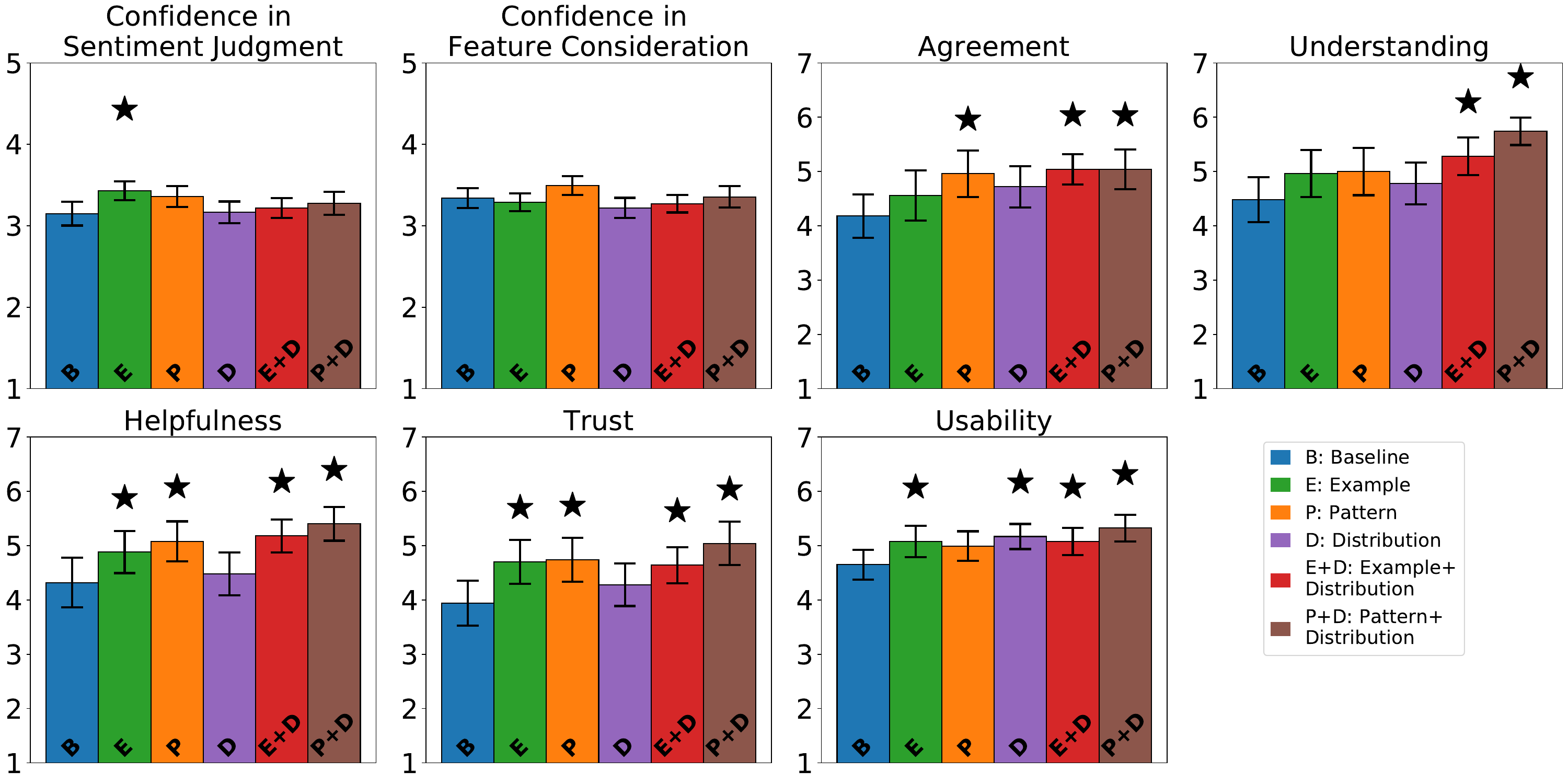}
\caption{Effects of different interface conditions on participants' \textbf{perceptions} with means and 95\% confidence intervals. The star mark highlights interface conditions with statistical significance ($p < .05$) compared to the \textbf{\textsc{Baseline}} condition.}
\label{fig:perceptions}
\end{figure*}

In RQ2, we investigated the effects of different interface conditions on participants' perceptions of the provided tools and their experiences. Figure~\ref{fig:perceptions} shows our RQ2 results. The interface condition did not have significant effects on any workload measures. Thus, the corresponding plots are omitted. Our results found four main trends.

First, none of our interface conditions helped participants achieve higher confidence in sentiment judgment and feature consideration in general. Even though participants were able to make more correct judgments in certain interface conditions, their confidence was close to the midpoint (i.e., moderately confident) in all cases. This result implies that our tools did not significantly increase participants' confidence when making these judgments.

Second, compared to the \textsc{Baseline} condition, the \textsc{Distribution} condition did not significantly improve participants' perceptions. Interestingly, participants made significantly more correct judgments in the \textsc{Distribution} condition (RQ1 results). This contrast suggests that while the label distribution could persuade participants to make objectively correct judgments, it was not subjectively perceived as understandable, helpful, or trustworthy.

Third, the \textsc{Example+Distribution} and \textsc{Pattern+Distribution} conditions significantly improved participants' perceptions across all measures except the confidence measure. These conditions with two tools might provide a more comprehensive view of word-sentiment relations than conditions with only one tool. It also demonstrates the necessity of providing more concrete explanations in addition to only showing the label distribution.

Finally, none of the interface conditions had effects on workload measures compared to the \textsc{Baseline} condition. This implies that our tools improved participants' performance without increasing workload.

\subsection{RQ3: Behaviors}
\begin{figure*}[htbp]
\centering
\includegraphics[width=\textwidth]{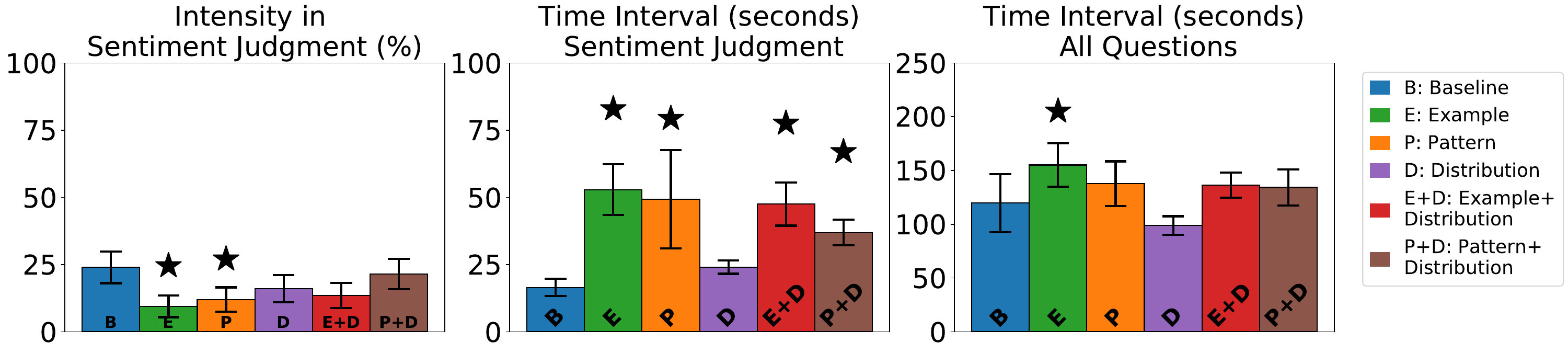}
\caption{Effects of different interface conditions on participants' \textbf{behaviors} with means and 95\% confidence intervals. The star mark highlights interface conditions with statistical significance ($p < .05$) compared to the \textbf{\textsc{Baseline}} condition.}
\label{fig:behaviors}
\end{figure*}

In RQ3, we investigated the effects of different interface conditions on participants' behaviors during different tasks. Figure~\ref{fig:behaviors} shows our RQ3 results. Our results found two main trends.

First, across all interface conditions, participants made intense sentiment judgments (i.e., selecting the two endpoints on the range slider) 25\% of the time or less. Participants made significantly less intense sentiment judgments in the \textsc{Example} and \textsc{Pattern} conditions. One possible reason is that participants received an equal or similar volume of explanations for both sentiments, ultimately leading to judgments of a more moderate intensity. In contrast, access to the label distribution made participants more likely to make an intense sentiment judgment.

Second, participants spent significantly more time on sentiment judgments in all interface conditions except the \textsc{Distribution} condition. This result is not surprising---the \textsc{Distribution} condition was simple as it only showed a pie chart of the sentiment label distribution. In contrast, other interface conditions provided more nuanced textual explanations and interactive functions, so participants engaged in activities such as reading and exploring the system. Regarding the time interval for the entire judgment process, only the \textsc{Example} condition significantly slowed participants down. One possible reason is that participants spent time reading detailed reviews before making judgments.

%% file: tex/discussion.tex
\section{Discussion}
In this section, we summarize the effects of our three tools, report on additional analyses regarding RQ1, discuss design implications, and review limitations of our study.

\textbf{Summary of Results:} In our study, we designed three tools to help participants understand the predictiveness of an unintuitive feature in a sentiment classifier. Our tools had different effects compared to the no-tool baseline. The \textsc{Distribution} tool, which showed the label distribution of training instances containing the word, represents the most abstract evidence. While it helped participants make sentiment and feature consideration judgments correctly and quickly, it did not improve perceptions. The \textsc{Example} tool, which provided a set of  training examples containing the word, is a natural approach to explaining unintuitive features. Using this tool, participants spent significantly longer time pondering on each question but failed to make more correct sentiment judgments. The \textsc{Pattern} tool, which extracted contextual patterns containing the word, provided a summary of underlying phenomena associated with the unintuitive feature. While it helped participants make more correct sentiment judgments and listed more correct scenarios, its effects on feature consideration were only marginally significant ($p=.056$). 

To summarize, no tool alone could help participants both: (1) understand the predictiveness of an unintuitive feature and (2) have better perceptions of the system and their experience.  The best approach is to provide a combination of tools (i.e., \textsc{Pattern+Distribution} and \textsc{Example+Distribution}).  Prior studies have found a similar trend.  That is, providing a combination of tools or visualizations helps end-users better understand a machine learning model than providing a single tool/visualization~\cite{gonzalez2020human,qu2021study,bussone2015role}.

\textbf{Additional RQ1 Analysis:} The listed scenarios in RQ1 reflected how participants rationalized unintuitive word-sentiment relations in their own words. They provide insights into different strategies participants took to explain unintuitive features. We analyzed these data regardless of interface conditions and  discovered five strategies. The dominant strategy was to list concrete examples that carry the sentiment and contain the word ($N = 3,030$ out of all $3,447$ listed scenarios). We report on four other strategies as follows.

First, participants explained the predictiveness of a word by describing \emph{semantic} patterns ($N = 347$), including a word's meaning (\textit{``\underline{heavy} means sturdy.''}), its typical usage in product reviews (\textit{``\underline{wait} indicates a slow shipping speed.''}), and its typical usage in natural language (\textit{``\underline{stay} is used for long-lasting and durable.''}).

Second, participants explained the predictiveness of a word by describing \emph{lexical} clues ($N = 11$).  This included explaining the use of an adjective or adverb (\textit{``\underline{completely} is an amplifier typically used for negative things.''}), the tense of a word (\textit{``\underline{lasted} implies that the product stopped working.''}), and the differences between a word's singular vs.~plural form (\textit{``\underline{star} is singular, thus less likely to be as positive [than] its plural form like `five stars'.''}).

Third, participants engaged in \emph{pragmatics} analysis ($N = 56$).  In such cases, participants thought about the intention and implication behind a word when used in a review. Examples included \textit{``people are more likely to complain about something special [...], especially something as small as a \underline{seal}.''} as well as \textit{``people don't tend to be happy if something just \underline{works}.''} Under the second and third strategies, participants described evidence that is highly nuanced.  From the lens of the dual process theory in psychology~\cite{kahneman2011thinking}, these participants exhibited more thoughtful and critical investigation, which is a different reasoning process compared to using heuristics such as a word's literal meaning.

Finally, some participants made judgments based solely on the label distribution ($N = 3$), (e.g., \textit{``I see \underline{zipper} is associated with negative reviews more.''}) Such scenarios were not considered valid. However, they demonstrate that some participants took shortcuts by referring to the majority label when explaining the predictiveness of a word.

\textbf{Design Implications:}  Our study provides two major implications for future designs of XAI tools.

First, \emph{XAI tools that provide feature importance explanations should prepare to further explain why a feature is predictive}.  In our study, we focused on predictive features in a relatively simple model (i.e., logistic regression using unigrams) and a simple task (i.e., sentiment classification). Our results found that many predictive features can be perceived as unintuitive to humans even in such a simple context.  The same is likely true for more complex models and tasks.  One possible explanation is that while machines learn predictive features from statistical patterns, humans understand a concept (e.g., sentiment) based on semantics, pragmatics, prior experience, and multi-step reasoning~\cite{sevastjanova2022beware,hu2022fine}. For example, ``minutes'' was predictive of \emph{positive} for automotive product reviews and \emph{negative} for pet product reviews.  Logistic regression models learned these statistical patterns without further asking \emph{why}.  However, for these trends to make sense to humans, it is helpful to further realize that ``minutes'' in automotive product reviews is typically used to describe a quick installation and that ``minutes'' in pet product reviews is typically used to describe a short product lifespan.  These additional explanations are helpful because they provide the broader context that humans need to position and reason about an otherwise unintuitive statistical pattern (i.e., ``feature $x$ is predictive of label $y$'').

Second, \emph{predictive yet unintuitive features can be attributed to relevant patterns and contexts in the training data}.
While prior studies employed pairwise feature interactions in a local example to explain predictive yet unintuitive unigrams ~\cite{tsang2018can,borisov2022relational,janizek2021explaining}, our tools explained unintuitive unigrams by tracing them back to the origin---the label distribution, relevant examples, and contextual patterns extracted from training data.  Our results confirm the efficacy of this approach.  Compared to participants without access to any tools (i.e., \textsc{Baseline} condition), participants who had access to our contextual pattern tool and distribution tool (i.e., \textsc{Pattern+Distribution} condition) were better able to correctly judge a word’s sentiment, correctly judge a word’s predictive power, describe scenarios for \emph{why} a word is predictive, and had better perceptions of the AI system and their experiences.  This finding suggests two important points supported by prior work.  First, end-users benefit from being able to scrutinize \emph{why} a feature is predictive in addition to knowing predictive features only~\cite{kulesza2015principles, kulesza2009fixing}.  Second, communicating information about the training data is an effective approach to enhancing an end-user’s understanding of and trust in a machine learning model~\cite{anik2021data}.

\textbf{Limitations and Opportunities for Future Work:} Our study has several limitations.  First, we focused on unintuitive features in the context of sentiment analysis.  Explaining unintuitive features in more complex NLP tasks where word-label relations are more nuanced such as deception detection~\cite{lai2020chicago}, toxicity detection~\cite{carton2020feature}, and sarcasm detection~\cite{joshi2017automatic} may require new tools.  Second, we focused on explaining unintuitive unigrams learned by a logistic regression classifier.  Future work should explore the generalizability of our tools to other feature representations and models (e.g., non-linear models).  Finally, in our study, participants were asked to scrutinize individual features that were predicted to be unintuitive.  Future work could explore visualizations that nudge users to recognize that a feature is unintuitive and explore such unintuitive features based on their own curiosity.

%% file: tex/conclusion.tex
\section{Conclusion}
Although showing predictive features is a prevalent approach to explaining machine learning models, features deemed as predictive by machines can be incomprehensible or unintuitive 
to humans. Predictive yet unintuitive features often represent phenomena that are not obvious without additional explanations. 
Our research took initial steps toward explaining unintuitive features by using sentiment analysis of product reviews as a case study. We focused on explaining the predictiveness of an unintuitive unigram feature by showing (1) label distribution, (2) sampled training examples, and (3) contextual patterns mined from training data. We conducted a crowdsourced study ($N=300$) to evaluate the efficacy of our tools from different perspectives. 
While the quantitative label distribution could quickly convince  participants to accept the unintuitive association between a feature and a label, seeing concrete examples and especially contextual patterns improved participants' qualitative understanding of the underlying phenomena and subjective perceptions of the provided tools. The best results were achieved when the tools where combined. 
Our research shows that it is both \emph{possible} and \emph{useful} to  explain predictive yet unintuitive features learned by sentiment classifiers. It opens up research opportunities to investigating problems of similar nature in a wider range of tasks and models.

%% file: tex/ethical_stmt.tex
\section*{Ethics Statements}

\textbf{Ethical considerations statement:} The study was reviewed and approved by our Institutional Review Board (IRB). During the study, participants were asked to complete several tasks that were not considered to be too cognitively demanding. The study used a between-subjects design.  This was partly done to help prevent a ``spill-over effect'' between interface conditions within a single participant. 

\textbf{Researcher positionality statement:} This work was partly inspired by our experiences as instructors of graduate-level text data mining courses, where curious students are frequently puzzled by certain features that are learned to be important according a model but do not make immediate sense to humans. We found that such questions cannot be sufficiently answered by current explainable machine learning techniques. We therefore set out to develop and evaluate tools to bridge this gap, hoping these tool can help people learn about a predictive task and increase their understanding of and trust in machine learning models.

\textbf{Adverse impact statement:} In our study, we developed different tools to help people understand the predictiveness of an unintuitive feature.  Our tools focused on displaying quantitative evidence (e.g., the label distribution across training instances containing the word) and qualitative evidence (e.g., contextual patterns containing the word in the training data).  Our RQ1 results suggest that quantitative evidence alone can influence people to accept that a feature is strongly predictive without understanding \emph{why}.  As shown in Figure~\ref{fig:understanding}, participants in the \textsc{Distribution} (vs.~\textsc{Baseline} condition) were significantly more likely to accept that a feature is strongly predictive but were \emph{not} significantly more likely to list valid scenarios to justify their judgment.  This result underscores the importance of showing both quantitative and qualitative evidence.  Showing only quantitative evidence may influence people to superficially accept that a machine learning model is doing the ``right thing'' without critically understanding why.

%% file: tex/appendix.tex
\appendix

\section{Study Task Allocation}\label{appendix_a}

\begin{figure*}[htbp]
\centering
\includegraphics[width=\textwidth]{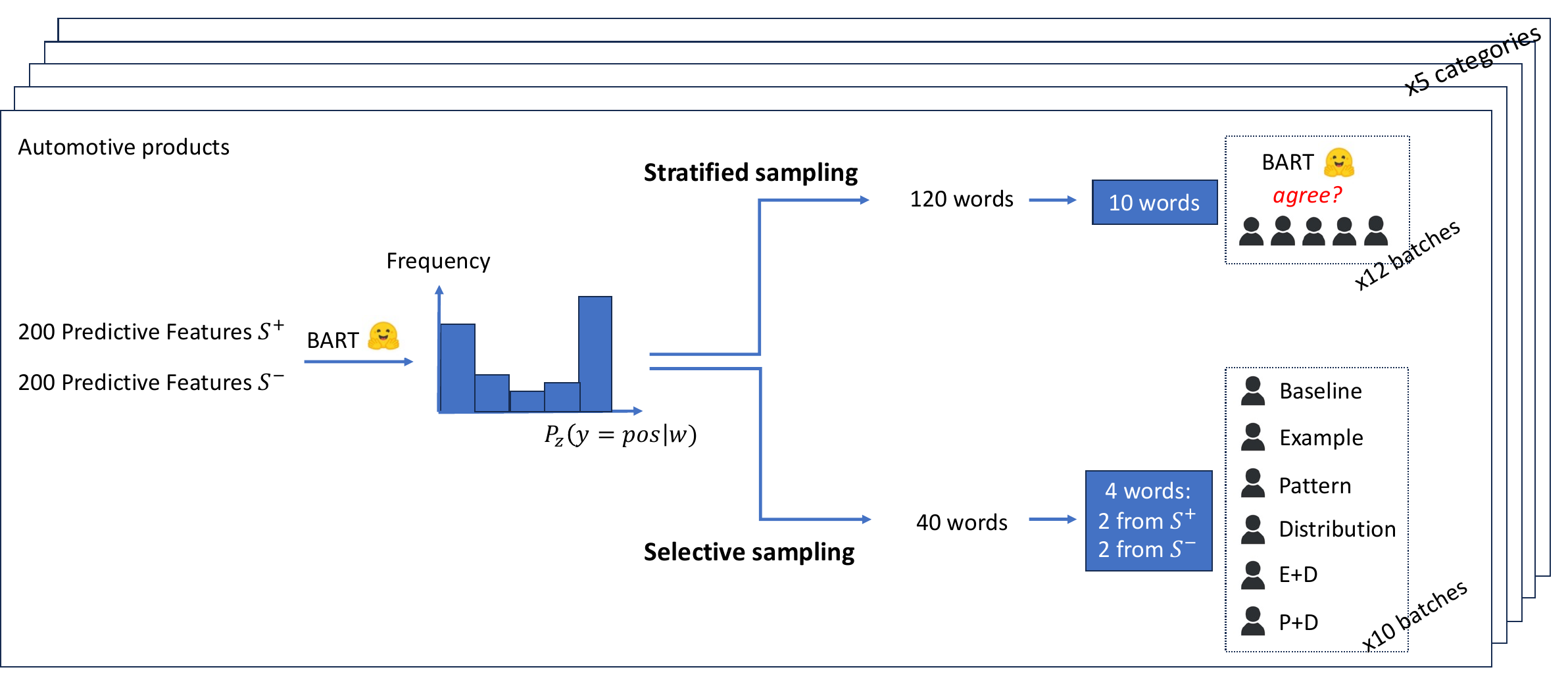}
\caption{Word sampling and task allocation. We used the zero-shot classifier to estimate $P_z(y=pos|w)$, the probability that word $w$ conveys positive sentiment, for every word in sets $\mathcal{S}^+$ and $\mathcal{S}^-$ (i.e., predictive of positive and negative according to the logistic regression model). For \textsc{Phase 1}, we applied stratified sampling to select 120 words and organized them into 12 batches of 10 words each. Each batch was redundantly judged by five participants. For \textsc{Phase 2}, we applied selective sampling to select 40 words and organized them into 10 batches of 4 words each. Each batch was redundantly judged by six participants, each in a different interface condition. Three criteria were used for the selective sampling. First, we included words with $P_z(y=pos|w) < 0.2$ from set $\mathcal{S}^+$.  These are words that are \emph{paradoxically} predictive of positive. Second, we included words with $P_z(y=pos|w) > 0.8$ from set $\mathcal{S}^-$.  These are words that are \emph{paradoxically} predictive of negative.  Third, we included words with $0.2 \le P_z(y=pos|w) \le 0.8$ from sets $\mathcal{S}^+$ and $\mathcal{S}^-$.  These are words that are unintuitive regardless of which sentiment they are predictive of. }
\end{figure*}

\newpage
\section{\textsc{Phase 2} System Interface}\label{appendix_b}

\begin{figure*}[htbp]
\centering
\includegraphics[width=\textwidth]{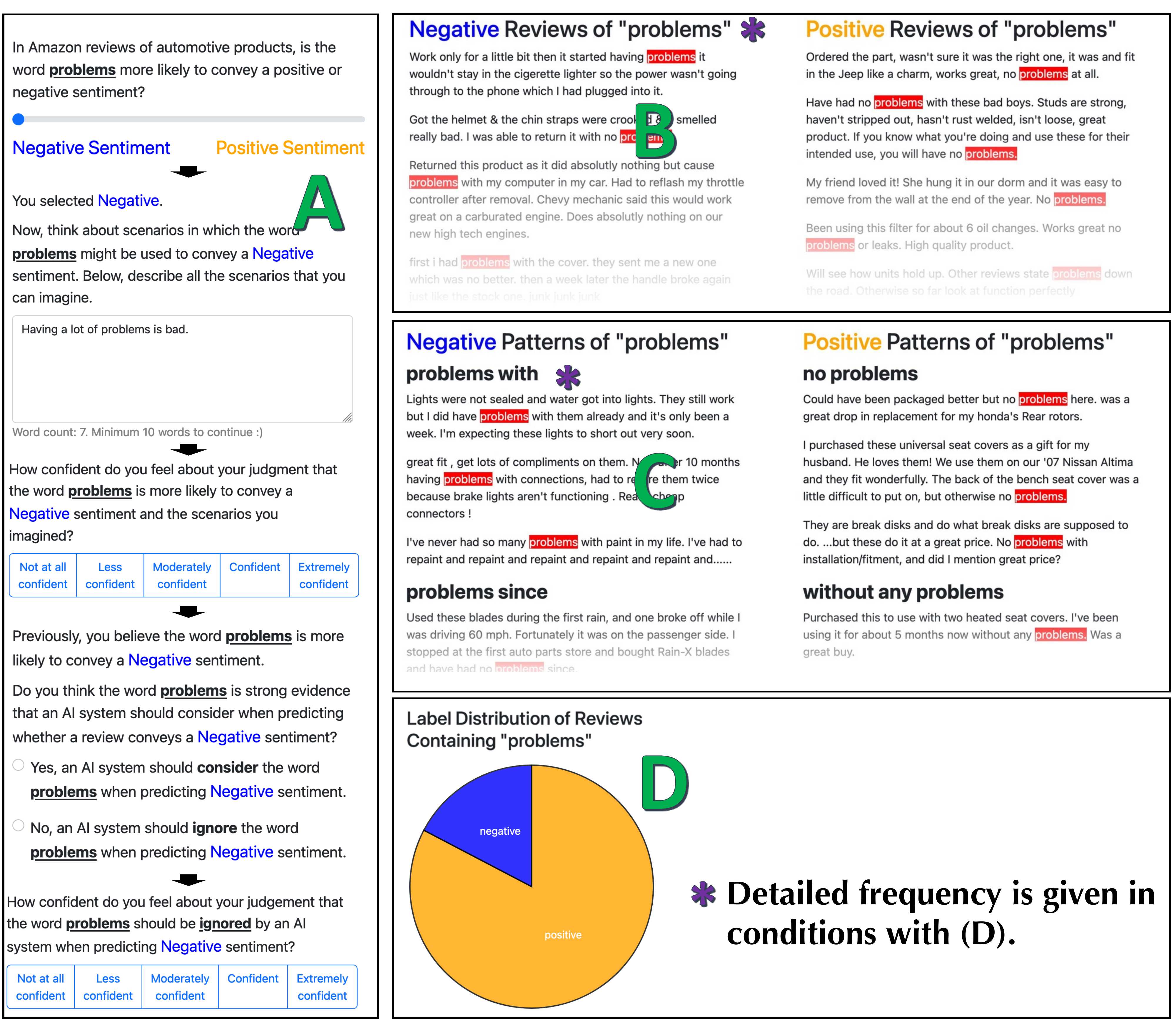}
\caption{\textsc{Phase 2} questions and tools. In the \textsc{Baseline} condition, question box (A) was positioned in the middle. Otherwise, it was positioned on the left side. Questions were displayed on consecutive pages instead of simultaneously on one page. Tools were presented on the right side: (B)-\textbf{\textsc{Example}}, (C)-\textbf{\textsc{Pattern}} and (D)-\textbf{\textsc{Distribution}}.}
\end{figure*}

%% file: main.bbl

\begin{thebibliography}{55}


\ifx \showCODEN    \undefined \def \showCODEN     #1{\unskip}     \fi
\ifx \showDOI      \undefined \def \showDOI       #1{#1}\fi
\ifx \showISBNx    \undefined \def \showISBNx     #1{\unskip}     \fi
\ifx \showISBNxiii \undefined \def \showISBNxiii  #1{\unskip}     \fi
\ifx \showISSN     \undefined \def \showISSN      #1{\unskip}     \fi
\ifx \showLCCN     \undefined \def \showLCCN      #1{\unskip}     \fi
\ifx \shownote     \undefined \def \shownote      #1{#1}          \fi
\ifx \showarticletitle \undefined \def \showarticletitle #1{#1}   \fi
\ifx \showURL      \undefined \def \showURL       {\relax}        \fi
\providecommand\bibfield[2]{#2}
\providecommand\bibinfo[2]{#2}
\providecommand\natexlab[1]{#1}
\providecommand\showeprint[2][]{arXiv:#2}

\bibitem[\protect\citeauthoryear{Anik and Bunt}{Anik and Bunt}{2021}]%
        {anik2021data}
\bibfield{author}{\bibinfo{person}{Ariful~Islam Anik} {and}
  \bibinfo{person}{Andrea Bunt}.} \bibinfo{year}{2021}\natexlab{}.
\newblock \showarticletitle{Data-centric explanations: explaining training data
  of machine learning systems to promote transparency}. In
  \bibinfo{booktitle}{\emph{Proceedings of the 2021 CHI Conference on Human
  Factors in Computing Systems}}. \bibinfo{pages}{1--13}.
\newblock


\bibitem[\protect\citeauthoryear{Bansal, Wu, Zhou, Fok, Nushi, Kamar, Ribeiro,
  and Weld}{Bansal et~al\mbox{.}}{2021}]%
        {bansal2021does}
\bibfield{author}{\bibinfo{person}{Gagan Bansal}, \bibinfo{person}{Tongshuang
  Wu}, \bibinfo{person}{Joyce Zhou}, \bibinfo{person}{Raymond Fok},
  \bibinfo{person}{Besmira Nushi}, \bibinfo{person}{Ece Kamar},
  \bibinfo{person}{Marco~Tulio Ribeiro}, {and} \bibinfo{person}{Daniel Weld}.}
  \bibinfo{year}{2021}\natexlab{}.
\newblock \showarticletitle{Does the whole exceed its parts? the effect of ai
  explanations on complementary team performance}. In
  \bibinfo{booktitle}{\emph{Proceedings of the 2021 CHI Conference on Human
  Factors in Computing Systems}}. \bibinfo{pages}{1--16}.
\newblock


\bibitem[\protect\citeauthoryear{Bird, Klein, and Loper}{Bird
  et~al\mbox{.}}{2009}]%
        {bird2009natural}
\bibfield{author}{\bibinfo{person}{Steven Bird}, \bibinfo{person}{Ewan Klein},
  {and} \bibinfo{person}{Edward Loper}.} \bibinfo{year}{2009}\natexlab{}.
\newblock \bibinfo{booktitle}{\emph{Natural language processing with Python:
  analyzing text with the natural language toolkit}}.
\newblock \bibinfo{publisher}{" O'Reilly Media, Inc."}.
\newblock


\bibitem[\protect\citeauthoryear{Borisov and Kasneci}{Borisov and
  Kasneci}{2022}]%
        {borisov2022relational}
\bibfield{author}{\bibinfo{person}{Vadim Borisov} {and}
  \bibinfo{person}{Gjergji Kasneci}.} \bibinfo{year}{2022}\natexlab{}.
\newblock \showarticletitle{Relational Local Explanations}.
\newblock \bibinfo{journal}{\emph{arXiv preprint arXiv:2212.12374}}
  (\bibinfo{year}{2022}).
\newblock


\bibitem[\protect\citeauthoryear{Brooke}{Brooke}{2013}]%
        {brooke2013sus}
\bibfield{author}{\bibinfo{person}{John Brooke}.}
  \bibinfo{year}{2013}\natexlab{}.
\newblock \showarticletitle{SUS: a retrospective}.
\newblock \bibinfo{journal}{\emph{Journal of usability studies}}
  \bibinfo{volume}{8}, \bibinfo{number}{2} (\bibinfo{year}{2013}),
  \bibinfo{pages}{29--40}.
\newblock


\bibitem[\protect\citeauthoryear{Bu{\c{c}}inca, Lin, Gajos, and
  Glassman}{Bu{\c{c}}inca et~al\mbox{.}}{2020}]%
        {buccinca2020proxy}
\bibfield{author}{\bibinfo{person}{Zana Bu{\c{c}}inca}, \bibinfo{person}{Phoebe
  Lin}, \bibinfo{person}{Krzysztof~Z Gajos}, {and} \bibinfo{person}{Elena~L
  Glassman}.} \bibinfo{year}{2020}\natexlab{}.
\newblock \showarticletitle{Proxy tasks and subjective measures can be
  misleading in evaluating explainable AI systems}. In
  \bibinfo{booktitle}{\emph{Proceedings of the 25th international conference on
  intelligent user interfaces}}. \bibinfo{pages}{454--464}.
\newblock


\bibitem[\protect\citeauthoryear{Bursac, Gauss, Williams, and Hosmer}{Bursac
  et~al\mbox{.}}{2008}]%
        {bursac2008purposeful}
\bibfield{author}{\bibinfo{person}{Zoran Bursac}, \bibinfo{person}{C~Heath
  Gauss}, \bibinfo{person}{David~Keith Williams}, {and}
  \bibinfo{person}{David~W Hosmer}.} \bibinfo{year}{2008}\natexlab{}.
\newblock \showarticletitle{Purposeful selection of variables in logistic
  regression}.
\newblock \bibinfo{journal}{\emph{Source code for biology and medicine}}
  \bibinfo{volume}{3}, \bibinfo{number}{1} (\bibinfo{year}{2008}),
  \bibinfo{pages}{1--8}.
\newblock


\bibitem[\protect\citeauthoryear{Bussone, Stumpf, and O'Sullivan}{Bussone
  et~al\mbox{.}}{2015}]%
        {bussone2015role}
\bibfield{author}{\bibinfo{person}{Adrian Bussone}, \bibinfo{person}{Simone
  Stumpf}, {and} \bibinfo{person}{Dympna O'Sullivan}.}
  \bibinfo{year}{2015}\natexlab{}.
\newblock \showarticletitle{The role of explanations on trust and reliance in
  clinical decision support systems}. In \bibinfo{booktitle}{\emph{2015
  international conference on healthcare informatics}}. IEEE,
  \bibinfo{pages}{160--169}.
\newblock


\bibitem[\protect\citeauthoryear{Cai, Jongejan, and Holbrook}{Cai
  et~al\mbox{.}}{2019a}]%
        {cai2019effects}
\bibfield{author}{\bibinfo{person}{Carrie~J Cai}, \bibinfo{person}{Jonas
  Jongejan}, {and} \bibinfo{person}{Jess Holbrook}.}
  \bibinfo{year}{2019}\natexlab{a}.
\newblock \showarticletitle{The effects of example-based explanations in a
  machine learning interface}. In \bibinfo{booktitle}{\emph{Proceedings of the
  24th international conference on intelligent user interfaces}}.
  \bibinfo{pages}{258--262}.
\newblock


\bibitem[\protect\citeauthoryear{Cai, Reif, Hegde, Hipp, Kim, Smilkov,
  Wattenberg, Viegas, Corrado, Stumpe, et~al\mbox{.}}{Cai
  et~al\mbox{.}}{2019b}]%
        {cai2019human}
\bibfield{author}{\bibinfo{person}{Carrie~J Cai}, \bibinfo{person}{Emily Reif},
  \bibinfo{person}{Narayan Hegde}, \bibinfo{person}{Jason Hipp},
  \bibinfo{person}{Been Kim}, \bibinfo{person}{Daniel Smilkov},
  \bibinfo{person}{Martin Wattenberg}, \bibinfo{person}{Fernanda Viegas},
  \bibinfo{person}{Greg~S Corrado}, \bibinfo{person}{Martin~C Stumpe},
  {et~al\mbox{.}}} \bibinfo{year}{2019}\natexlab{b}.
\newblock \showarticletitle{Human-centered tools for coping with imperfect
  algorithms during medical decision-making}. In
  \bibinfo{booktitle}{\emph{Proceedings of the 2019 CHI conference on human
  factors in computing systems}}. \bibinfo{pages}{1--14}.
\newblock


\bibitem[\protect\citeauthoryear{Carbonell and Goldstein}{Carbonell and
  Goldstein}{1998}]%
        {carbonell1998use}
\bibfield{author}{\bibinfo{person}{Jaime Carbonell} {and} \bibinfo{person}{Jade
  Goldstein}.} \bibinfo{year}{1998}\natexlab{}.
\newblock \showarticletitle{The use of MMR, diversity-based reranking for
  reordering documents and producing summaries}. In
  \bibinfo{booktitle}{\emph{Proceedings of the 21st annual international ACM
  SIGIR conference on Research and development in information retrieval}}.
  \bibinfo{pages}{335--336}.
\newblock


\bibitem[\protect\citeauthoryear{Carton, Mei, and Resnick}{Carton
  et~al\mbox{.}}{2020}]%
        {carton2020feature}
\bibfield{author}{\bibinfo{person}{Samuel Carton}, \bibinfo{person}{Qiaozhu
  Mei}, {and} \bibinfo{person}{Paul Resnick}.} \bibinfo{year}{2020}\natexlab{}.
\newblock \showarticletitle{Feature-Based Explanations Don't Help People Detect
  Misclassifications of Online Toxicity}. In
  \bibinfo{booktitle}{\emph{Proceedings of the International AAAI Conference on
  Web and Social Media}}, Vol.~\bibinfo{volume}{14}. \bibinfo{pages}{95--106}.
\newblock


\bibitem[\protect\citeauthoryear{Caruana, Lou, Gehrke, Koch, Sturm, and
  Elhadad}{Caruana et~al\mbox{.}}{2015}]%
        {caruana2015intelligible}
\bibfield{author}{\bibinfo{person}{Rich Caruana}, \bibinfo{person}{Yin Lou},
  \bibinfo{person}{Johannes Gehrke}, \bibinfo{person}{Paul Koch},
  \bibinfo{person}{Marc Sturm}, {and} \bibinfo{person}{Noemie Elhadad}.}
  \bibinfo{year}{2015}\natexlab{}.
\newblock \showarticletitle{Intelligible models for healthcare: Predicting
  pneumonia risk and hospital 30-day readmission}. In
  \bibinfo{booktitle}{\emph{Proceedings of the 21th ACM SIGKDD international
  conference on knowledge discovery and data mining}}.
  \bibinfo{pages}{1721--1730}.
\newblock


\bibitem[\protect\citeauthoryear{Chen, Liao, Wortman~Vaughan, and Bansal}{Chen
  et~al\mbox{.}}{2023}]%
        {chen2023understanding}
\bibfield{author}{\bibinfo{person}{Valerie Chen}, \bibinfo{person}{Q~Vera
  Liao}, \bibinfo{person}{Jennifer Wortman~Vaughan}, {and}
  \bibinfo{person}{Gagan Bansal}.} \bibinfo{year}{2023}\natexlab{}.
\newblock \showarticletitle{Understanding the role of human intuition on
  reliance in human-AI decision-making with explanations}.
\newblock \bibinfo{journal}{\emph{Proceedings of the ACM on Human-computer
  Interaction}} \bibinfo{volume}{7}, \bibinfo{number}{CSCW2}
  (\bibinfo{year}{2023}), \bibinfo{pages}{1--32}.
\newblock


\bibitem[\protect\citeauthoryear{Chromik, Eiband, Buchner, Kr{\"u}ger, and
  Butz}{Chromik et~al\mbox{.}}{2021}]%
        {chromik2021think}
\bibfield{author}{\bibinfo{person}{Michael Chromik}, \bibinfo{person}{Malin
  Eiband}, \bibinfo{person}{Felicitas Buchner}, \bibinfo{person}{Adrian
  Kr{\"u}ger}, {and} \bibinfo{person}{Andreas Butz}.}
  \bibinfo{year}{2021}\natexlab{}.
\newblock \showarticletitle{I think i get your point, AI! the illusion of
  explanatory depth in explainable AI}. In \bibinfo{booktitle}{\emph{26th
  International Conference on Intelligent User Interfaces}}.
  \bibinfo{pages}{307--317}.
\newblock


\bibitem[\protect\citeauthoryear{Covert, Lundberg, and Lee}{Covert
  et~al\mbox{.}}{2020}]%
        {covert2020understanding}
\bibfield{author}{\bibinfo{person}{Ian Covert}, \bibinfo{person}{Scott~M
  Lundberg}, {and} \bibinfo{person}{Su-In Lee}.}
  \bibinfo{year}{2020}\natexlab{}.
\newblock \showarticletitle{Understanding global feature contributions with
  additive importance measures}.
\newblock \bibinfo{journal}{\emph{Advances in Neural Information Processing
  Systems}}  \bibinfo{volume}{33} (\bibinfo{year}{2020}),
  \bibinfo{pages}{17212--17223}.
\newblock


\bibitem[\protect\citeauthoryear{Doshi-Velez and Kim}{Doshi-Velez and
  Kim}{2017}]%
        {doshi2017towards}
\bibfield{author}{\bibinfo{person}{Finale Doshi-Velez} {and}
  \bibinfo{person}{Been Kim}.} \bibinfo{year}{2017}\natexlab{}.
\newblock \showarticletitle{Towards A Rigorous Science of Interpretable Machine
  Learning}.
\newblock \bibinfo{journal}{\emph{arXiv}} (\bibinfo{year}{2017}).
\newblock
\urldef\tempurl%
\url{https://arxiv.org/abs/1702.08608}
\showURL{%
\tempurl}


\bibitem[\protect\citeauthoryear{Farrar and Glauber}{Farrar and
  Glauber}{1967}]%
        {farrar1967multicollinearity}
\bibfield{author}{\bibinfo{person}{Donald~E Farrar} {and}
  \bibinfo{person}{Robert~R Glauber}.} \bibinfo{year}{1967}\natexlab{}.
\newblock \showarticletitle{Multicollinearity in regression analysis: the
  problem revisited}.
\newblock \bibinfo{journal}{\emph{The Review of Economic and Statistics}}
  (\bibinfo{year}{1967}), \bibinfo{pages}{92--107}.
\newblock


\bibitem[\protect\citeauthoryear{Gonzalez, Bansal, Fan, Jia, Mehdad, and
  Iyer}{Gonzalez et~al\mbox{.}}{2020}]%
        {gonzalez2020human}
\bibfield{author}{\bibinfo{person}{Ana~Valeria Gonzalez},
  \bibinfo{person}{Gagan Bansal}, \bibinfo{person}{Angela Fan},
  \bibinfo{person}{Robin Jia}, \bibinfo{person}{Yashar Mehdad}, {and}
  \bibinfo{person}{Srinivasan Iyer}.} \bibinfo{year}{2020}\natexlab{}.
\newblock \showarticletitle{Human evaluation of spoken vs. visual explanations
  for open-domain qa}.
\newblock \bibinfo{journal}{\emph{arXiv preprint arXiv:2012.15075}}
  (\bibinfo{year}{2020}).
\newblock


\bibitem[\protect\citeauthoryear{Hart and Staveland}{Hart and
  Staveland}{1988}]%
        {hart1988development}
\bibfield{author}{\bibinfo{person}{Sandra~G Hart} {and}
  \bibinfo{person}{Lowell~E Staveland}.} \bibinfo{year}{1988}\natexlab{}.
\newblock \showarticletitle{Development of NASA-TLX (Task Load Index): Results
  of empirical and theoretical research}.
\newblock In \bibinfo{booktitle}{\emph{Advances in psychology}}.
  Vol.~\bibinfo{volume}{52}. \bibinfo{publisher}{Elsevier},
  \bibinfo{pages}{139--183}.
\newblock


\bibitem[\protect\citeauthoryear{Hu, Floyd, Jouravlev, Fedorenko, and
  Gibson}{Hu et~al\mbox{.}}{2022}]%
        {hu2022fine}
\bibfield{author}{\bibinfo{person}{Jennifer Hu}, \bibinfo{person}{Sammy Floyd},
  \bibinfo{person}{Olessia Jouravlev}, \bibinfo{person}{Evelina Fedorenko},
  {and} \bibinfo{person}{Edward Gibson}.} \bibinfo{year}{2022}\natexlab{}.
\newblock \showarticletitle{A fine-grained comparison of pragmatic language
  understanding in humans and language models}.
\newblock \bibinfo{journal}{\emph{arXiv preprint arXiv:2212.06801}}
  (\bibinfo{year}{2022}).
\newblock


\bibitem[\protect\citeauthoryear{Janizek, Sturmfels, and Lee}{Janizek
  et~al\mbox{.}}{2021}]%
        {janizek2021explaining}
\bibfield{author}{\bibinfo{person}{Joseph~D Janizek}, \bibinfo{person}{Pascal
  Sturmfels}, {and} \bibinfo{person}{Su-In Lee}.}
  \bibinfo{year}{2021}\natexlab{}.
\newblock \showarticletitle{Explaining Explanations: Axiomatic Feature
  Interactions for Deep Networks.}
\newblock \bibinfo{journal}{\emph{J. Mach. Learn. Res.}}  \bibinfo{volume}{22}
  (\bibinfo{year}{2021}), \bibinfo{pages}{104--1}.
\newblock


\bibitem[\protect\citeauthoryear{Joshi, Bhattacharyya, and Carman}{Joshi
  et~al\mbox{.}}{2017}]%
        {joshi2017automatic}
\bibfield{author}{\bibinfo{person}{Aditya Joshi}, \bibinfo{person}{Pushpak
  Bhattacharyya}, {and} \bibinfo{person}{Mark~J Carman}.}
  \bibinfo{year}{2017}\natexlab{}.
\newblock \showarticletitle{Automatic sarcasm detection: A survey}.
\newblock \bibinfo{journal}{\emph{ACM Computing Surveys (CSUR)}}
  \bibinfo{volume}{50}, \bibinfo{number}{5} (\bibinfo{year}{2017}),
  \bibinfo{pages}{1--22}.
\newblock


\bibitem[\protect\citeauthoryear{Kahneman}{Kahneman}{2011}]%
        {kahneman2011thinking}
\bibfield{author}{\bibinfo{person}{Daniel Kahneman}.}
  \bibinfo{year}{2011}\natexlab{}.
\newblock \bibinfo{booktitle}{\emph{Thinking, fast and slow}}.
\newblock \bibinfo{publisher}{macmillan}.
\newblock


\bibitem[\protect\citeauthoryear{Kaufmann and Beehr}{Kaufmann and
  Beehr}{1986}]%
        {kaufmann1986interactions}
\bibfield{author}{\bibinfo{person}{Gary~M Kaufmann} {and}
  \bibinfo{person}{Terry~A Beehr}.} \bibinfo{year}{1986}\natexlab{}.
\newblock \showarticletitle{Interactions between job stressors and social
  support: Some counterintuitive results.}
\newblock \bibinfo{journal}{\emph{Journal of applied psychology}}
  \bibinfo{volume}{71}, \bibinfo{number}{3} (\bibinfo{year}{1986}),
  \bibinfo{pages}{522}.
\newblock


\bibitem[\protect\citeauthoryear{Kennedy}{Kennedy}{2005}]%
        {kennedy2005oh}
\bibfield{author}{\bibinfo{person}{Peter~E Kennedy}.}
  \bibinfo{year}{2005}\natexlab{}.
\newblock \showarticletitle{Oh no! I got the wrong sign! What should I do?}
\newblock \bibinfo{journal}{\emph{The Journal of Economic Education}}
  \bibinfo{volume}{36}, \bibinfo{number}{1} (\bibinfo{year}{2005}),
  \bibinfo{pages}{77--92}.
\newblock


\bibitem[\protect\citeauthoryear{Koh and Liang}{Koh and Liang}{2017}]%
        {koh2017understanding}
\bibfield{author}{\bibinfo{person}{Pang~Wei Koh} {and} \bibinfo{person}{Percy
  Liang}.} \bibinfo{year}{2017}\natexlab{}.
\newblock \showarticletitle{Understanding black-box predictions via influence
  functions}. In \bibinfo{booktitle}{\emph{International conference on machine
  learning}}. PMLR, \bibinfo{pages}{1885--1894}.
\newblock


\bibitem[\protect\citeauthoryear{Kulesza, Burnett, Wong, and Stumpf}{Kulesza
  et~al\mbox{.}}{2015}]%
        {kulesza2015principles}
\bibfield{author}{\bibinfo{person}{Todd Kulesza}, \bibinfo{person}{Margaret
  Burnett}, \bibinfo{person}{Weng-Keen Wong}, {and} \bibinfo{person}{Simone
  Stumpf}.} \bibinfo{year}{2015}\natexlab{}.
\newblock \showarticletitle{Principles of explanatory debugging to personalize
  interactive machine learning}. In \bibinfo{booktitle}{\emph{Proceedings of
  the 20th international conference on intelligent user interfaces}}.
  \bibinfo{pages}{126--137}.
\newblock


\bibitem[\protect\citeauthoryear{Kulesza, Wong, Stumpf, Perona, White, Burnett,
  Oberst, and Ko}{Kulesza et~al\mbox{.}}{2009}]%
        {kulesza2009fixing}
\bibfield{author}{\bibinfo{person}{Todd Kulesza}, \bibinfo{person}{Weng-Keen
  Wong}, \bibinfo{person}{Simone Stumpf}, \bibinfo{person}{Stephen Perona},
  \bibinfo{person}{Rachel White}, \bibinfo{person}{Margaret~M Burnett},
  \bibinfo{person}{Ian Oberst}, {and} \bibinfo{person}{Amy~J Ko}.}
  \bibinfo{year}{2009}\natexlab{}.
\newblock \showarticletitle{Fixing the program my computer learned: Barriers
  for end users, challenges for the machine}. In
  \bibinfo{booktitle}{\emph{Proceedings of the 14th international conference on
  Intelligent user interfaces}}. \bibinfo{pages}{187--196}.
\newblock


\bibitem[\protect\citeauthoryear{Lai, Liu, and Tan}{Lai et~al\mbox{.}}{2020}]%
        {lai2020chicago}
\bibfield{author}{\bibinfo{person}{Vivian Lai}, \bibinfo{person}{Han Liu},
  {and} \bibinfo{person}{Chenhao Tan}.} \bibinfo{year}{2020}\natexlab{}.
\newblock \showarticletitle{" Why is' Chicago'deceptive?" Towards Building
  Model-Driven Tutorials for Humans}. In \bibinfo{booktitle}{\emph{Proceedings
  of the 2020 CHI Conference on Human Factors in Computing Systems}}.
  \bibinfo{pages}{1--13}.
\newblock


\bibitem[\protect\citeauthoryear{Lai and Tan}{Lai and Tan}{2019}]%
        {lai2019human}
\bibfield{author}{\bibinfo{person}{Vivian Lai} {and} \bibinfo{person}{Chenhao
  Tan}.} \bibinfo{year}{2019}\natexlab{}.
\newblock \showarticletitle{On human predictions with explanations and
  predictions of machine learning models: A case study on deception detection}.
  In \bibinfo{booktitle}{\emph{Proceedings of the conference on fairness,
  accountability, and transparency}}. \bibinfo{pages}{29--38}.
\newblock


\bibitem[\protect\citeauthoryear{Landis and Koch}{Landis and Koch}{1977}]%
        {landis1977agreement}
\bibfield{author}{\bibinfo{person}{J.~R. Landis} {and} \bibinfo{person}{G.~G.
  Koch}.} \bibinfo{year}{1977}\natexlab{}.
\newblock \showarticletitle{The Measurement of Observer Agreement for
  Categorical Data}.
\newblock \bibinfo{journal}{\emph{Biometrics}} \bibinfo{volume}{33},
  \bibinfo{number}{1} (\bibinfo{year}{1977}), \bibinfo{pages}{159--174}.
\newblock


\bibitem[\protect\citeauthoryear{Lewis, Liu, Goyal, Ghazvininejad, Mohamed,
  Levy, Stoyanov, and Zettlemoyer}{Lewis et~al\mbox{.}}{2020}]%
        {lewis2020bart}
\bibfield{author}{\bibinfo{person}{Mike Lewis}, \bibinfo{person}{Yinhan Liu},
  \bibinfo{person}{Naman Goyal}, \bibinfo{person}{Marjan Ghazvininejad},
  \bibinfo{person}{Abdelrahman Mohamed}, \bibinfo{person}{Omer Levy},
  \bibinfo{person}{Veselin Stoyanov}, {and} \bibinfo{person}{Luke
  Zettlemoyer}.} \bibinfo{year}{2020}\natexlab{}.
\newblock \showarticletitle{BART: Denoising Sequence-to-Sequence Pre-training
  for Natural Language Generation, Translation, and Comprehension}. In
  \bibinfo{booktitle}{\emph{Proceedings of the 58th Annual Meeting of the
  Association for Computational Linguistics}}. \bibinfo{pages}{7871--7880}.
\newblock


\bibitem[\protect\citeauthoryear{Lundberg and Lee}{Lundberg and Lee}{2017}]%
        {lundberg2017unified}
\bibfield{author}{\bibinfo{person}{Scott~M Lundberg} {and}
  \bibinfo{person}{Su-In Lee}.} \bibinfo{year}{2017}\natexlab{}.
\newblock \showarticletitle{A unified approach to interpreting model
  predictions}.
\newblock \bibinfo{journal}{\emph{Advances in neural information processing
  systems}}  \bibinfo{volume}{30} (\bibinfo{year}{2017}).
\newblock


\bibitem[\protect\citeauthoryear{Nathans, Oswald, and Nimon}{Nathans
  et~al\mbox{.}}{2012}]%
        {nathans2012interpreting}
\bibfield{author}{\bibinfo{person}{Laura~L Nathans},
  \bibinfo{person}{Frederick~L Oswald}, {and} \bibinfo{person}{Kim Nimon}.}
  \bibinfo{year}{2012}\natexlab{}.
\newblock \showarticletitle{Interpreting multiple linear regression: a
  guidebook of variable importance.}
\newblock \bibinfo{journal}{\emph{Practical assessment, research \&
  evaluation}} \bibinfo{volume}{17}, \bibinfo{number}{9}
  (\bibinfo{year}{2012}), \bibinfo{pages}{n9}.
\newblock


\bibitem[\protect\citeauthoryear{Nguyen}{Nguyen}{2018}]%
        {nguyen2018comparing}
\bibfield{author}{\bibinfo{person}{Dong Nguyen}.}
  \bibinfo{year}{2018}\natexlab{}.
\newblock \showarticletitle{Comparing automatic and human evaluation of local
  explanations for text classification}. In \bibinfo{booktitle}{\emph{16th
  Annual Conference of the North American Chapter of the Association for
  Computational Linguistics: Human Language Technologies}}. Association for
  Computational Linguistics, \bibinfo{pages}{1069--1078}.
\newblock


\bibitem[\protect\citeauthoryear{Ni, Li, and McAuley}{Ni et~al\mbox{.}}{2019}]%
        {ni2019justifying}
\bibfield{author}{\bibinfo{person}{Jianmo Ni}, \bibinfo{person}{Jiacheng Li},
  {and} \bibinfo{person}{Julian McAuley}.} \bibinfo{year}{2019}\natexlab{}.
\newblock \showarticletitle{Justifying recommendations using distantly-labeled
  reviews and fine-grained aspects}. In \bibinfo{booktitle}{\emph{Proceedings
  of the 2019 conference on empirical methods in natural language processing
  and the 9th international joint conference on natural language processing
  (EMNLP-IJCNLP)}}. \bibinfo{pages}{188--197}.
\newblock


\bibitem[\protect\citeauthoryear{Oh, Kim, Choi, Eun, Kim, Kim, Lee, and Suh}{Oh
  et~al\mbox{.}}{2020}]%
        {oh2020understanding}
\bibfield{author}{\bibinfo{person}{Changhoon Oh}, \bibinfo{person}{Seonghyeon
  Kim}, \bibinfo{person}{Jinhan Choi}, \bibinfo{person}{Jinsu Eun},
  \bibinfo{person}{Soomin Kim}, \bibinfo{person}{Juho Kim},
  \bibinfo{person}{Joonhwan Lee}, {and} \bibinfo{person}{Bongwon Suh}.}
  \bibinfo{year}{2020}\natexlab{}.
\newblock \showarticletitle{Understanding How People Reason about Aesthetic
  Evaluations of Artificial Intelligence}. In
  \bibinfo{booktitle}{\emph{Proceedings of the 2020 ACM Designing Interactive
  Systems Conference}}. \bibinfo{pages}{1169--1181}.
\newblock


\bibitem[\protect\citeauthoryear{Pedregosa, Varoquaux, Gramfort, Michel,
  Thirion, Grisel, Blondel, Prettenhofer, Weiss, Dubourg,
  et~al\mbox{.}}{Pedregosa et~al\mbox{.}}{2011}]%
        {pedregosa2011scikit}
\bibfield{author}{\bibinfo{person}{Fabian Pedregosa}, \bibinfo{person}{Ga{\"e}l
  Varoquaux}, \bibinfo{person}{Alexandre Gramfort}, \bibinfo{person}{Vincent
  Michel}, \bibinfo{person}{Bertrand Thirion}, \bibinfo{person}{Olivier
  Grisel}, \bibinfo{person}{Mathieu Blondel}, \bibinfo{person}{Peter
  Prettenhofer}, \bibinfo{person}{Ron Weiss}, \bibinfo{person}{Vincent
  Dubourg}, {et~al\mbox{.}}} \bibinfo{year}{2011}\natexlab{}.
\newblock \showarticletitle{Scikit-learn: Machine learning in Python}.
\newblock \bibinfo{journal}{\emph{the Journal of machine Learning research}}
  \bibinfo{volume}{12} (\bibinfo{year}{2011}), \bibinfo{pages}{2825--2830}.
\newblock


\bibitem[\protect\citeauthoryear{Pruthi, Liu, Kale, and Sundararajan}{Pruthi
  et~al\mbox{.}}{2020}]%
        {pruthi2020estimating}
\bibfield{author}{\bibinfo{person}{Garima Pruthi}, \bibinfo{person}{Frederick
  Liu}, \bibinfo{person}{Satyen Kale}, {and} \bibinfo{person}{Mukund
  Sundararajan}.} \bibinfo{year}{2020}\natexlab{}.
\newblock \showarticletitle{Estimating training data influence by tracing
  gradient descent}.
\newblock \bibinfo{journal}{\emph{Advances in Neural Information Processing
  Systems}}  \bibinfo{volume}{33} (\bibinfo{year}{2020}),
  \bibinfo{pages}{19920--19930}.
\newblock


\bibitem[\protect\citeauthoryear{Qu, Arguello, and Wang}{Qu
  et~al\mbox{.}}{2021}]%
        {qu2021study}
\bibfield{author}{\bibinfo{person}{Jiaming Qu}, \bibinfo{person}{Jaime
  Arguello}, {and} \bibinfo{person}{Yue Wang}.}
  \bibinfo{year}{2021}\natexlab{}.
\newblock \showarticletitle{A Study of Explainability Features to Scrutinize
  Faceted Filtering Results}. In \bibinfo{booktitle}{\emph{Proceedings of the
  30th ACM International Conference on Information \& Knowledge Management}}.
  \bibinfo{pages}{1498--1507}.
\newblock


\bibitem[\protect\citeauthoryear{Qu, Arguello, and Wang}{Qu
  et~al\mbox{.}}{2023}]%
        {qu2023}
\bibfield{author}{\bibinfo{person}{Jiaming Qu}, \bibinfo{person}{Jaime
  Arguello}, {and} \bibinfo{person}{Yue Wang}.}
  \bibinfo{year}{2023}\natexlab{}.
\newblock \showarticletitle{Understanding the Cognitive Influences of
  Interpretability Features on How Users Scrutinize Machine-Predicted
  Categories}. In \bibinfo{booktitle}{\emph{ACM SIGIR Conference On Human
  Information Interaction And Retrieval}}.
\newblock


\bibitem[\protect\citeauthoryear{Reimers and Gurevych}{Reimers and
  Gurevych}{2019}]%
        {reimers2019sentence}
\bibfield{author}{\bibinfo{person}{Nils Reimers} {and} \bibinfo{person}{Iryna
  Gurevych}.} \bibinfo{year}{2019}\natexlab{}.
\newblock \showarticletitle{Sentence-BERT: Sentence Embeddings using Siamese
  BERT-Networks}. In \bibinfo{booktitle}{\emph{Proceedings of the 2019
  Conference on Empirical Methods in Natural Language Processing and the 9th
  International Joint Conference on Natural Language Processing
  (EMNLP-IJCNLP)}}. \bibinfo{pages}{3982--3992}.
\newblock


\bibitem[\protect\citeauthoryear{Ribeiro and Contributors}{Ribeiro and
  Contributors}{2020}]%
        {lime2016github}
\bibfield{author}{\bibinfo{person}{Marco~Tulio Ribeiro} {and}
  \bibinfo{person}{GitHub Contributors}.} \bibinfo{year}{2020}\natexlab{}.
\newblock \bibinfo{booktitle}{\emph{{LIME: Explaining the predictions of any
  machine learning classifier}}}.
\newblock
\urldef\tempurl%
\url{https://github.com/marcotcr/lime}
\showURL{%
\tempurl}


\bibitem[\protect\citeauthoryear{Ribeiro, Singh, and Guestrin}{Ribeiro
  et~al\mbox{.}}{2016}]%
        {lime2016}
\bibfield{author}{\bibinfo{person}{Marco~Tulio Ribeiro},
  \bibinfo{person}{Sameer Singh}, {and} \bibinfo{person}{Carlos Guestrin}.}
  \bibinfo{year}{2016}\natexlab{}.
\newblock \showarticletitle{"Why should i trust you?" Explaining the
  predictions of any classifier}. In \bibinfo{booktitle}{\emph{Proceedings of
  the 22nd ACM SIGKDD international conference on knowledge discovery and data
  mining}}. \bibinfo{pages}{1135--1144}.
\newblock


\bibitem[\protect\citeauthoryear{Ribeiro, Singh, and Guestrin}{Ribeiro
  et~al\mbox{.}}{2018}]%
        {ribeiro2018anchors}
\bibfield{author}{\bibinfo{person}{Marco~Tulio Ribeiro},
  \bibinfo{person}{Sameer Singh}, {and} \bibinfo{person}{Carlos Guestrin}.}
  \bibinfo{year}{2018}\natexlab{}.
\newblock \showarticletitle{Anchors: High-precision model-agnostic
  explanations}. In \bibinfo{booktitle}{\emph{Proceedings of the AAAI
  conference on artificial intelligence}}, Vol.~\bibinfo{volume}{32}.
\newblock


\bibitem[\protect\citeauthoryear{Schuff, Jacovi, Adel, Goldberg, and Vu}{Schuff
  et~al\mbox{.}}{2022}]%
        {schuff2022human}
\bibfield{author}{\bibinfo{person}{Hendrik Schuff}, \bibinfo{person}{Alon
  Jacovi}, \bibinfo{person}{Heike Adel}, \bibinfo{person}{Yoav Goldberg}, {and}
  \bibinfo{person}{Ngoc~Thang Vu}.} \bibinfo{year}{2022}\natexlab{}.
\newblock \showarticletitle{Human interpretation of saliency-based explanation
  over text}. In \bibinfo{booktitle}{\emph{2022 ACM Conference on Fairness,
  Accountability, and Transparency}}. \bibinfo{pages}{611--636}.
\newblock


\bibitem[\protect\citeauthoryear{Sevastjanova and El-Assady}{Sevastjanova and
  El-Assady}{2022}]%
        {sevastjanova2022beware}
\bibfield{author}{\bibinfo{person}{Rita Sevastjanova} {and}
  \bibinfo{person}{Mennatallah El-Assady}.} \bibinfo{year}{2022}\natexlab{}.
\newblock \showarticletitle{Beware the rationalization trap! when language
  model explainability diverges from our mental models of language}.
\newblock \bibinfo{journal}{\emph{arXiv preprint arXiv:2207.06897}}
  (\bibinfo{year}{2022}).
\newblock


\bibitem[\protect\citeauthoryear{Shrikumar, Greenside, and Kundaje}{Shrikumar
  et~al\mbox{.}}{2017}]%
        {shrikumar2017learning}
\bibfield{author}{\bibinfo{person}{Avanti Shrikumar}, \bibinfo{person}{Peyton
  Greenside}, {and} \bibinfo{person}{Anshul Kundaje}.}
  \bibinfo{year}{2017}\natexlab{}.
\newblock \showarticletitle{Learning important features through propagating
  activation differences}. In \bibinfo{booktitle}{\emph{International
  conference on machine learning}}. PMLR, \bibinfo{pages}{3145--3153}.
\newblock


\bibitem[\protect\citeauthoryear{Sundararajan, Taly, and Yan}{Sundararajan
  et~al\mbox{.}}{2017}]%
        {sundararajan2017axiomatic}
\bibfield{author}{\bibinfo{person}{Mukund Sundararajan}, \bibinfo{person}{Ankur
  Taly}, {and} \bibinfo{person}{Qiqi Yan}.} \bibinfo{year}{2017}\natexlab{}.
\newblock \showarticletitle{Axiomatic attribution for deep networks}. In
  \bibinfo{booktitle}{\emph{International conference on machine learning}}.
  PMLR, \bibinfo{pages}{3319--3328}.
\newblock


\bibitem[\protect\citeauthoryear{Tsang, Sun, Ren, Xin, and Liu}{Tsang
  et~al\mbox{.}}{2018}]%
        {tsang2018can}
\bibfield{author}{\bibinfo{person}{Michael Tsang}, \bibinfo{person}{Youbang
  Sun}, \bibinfo{person}{Dongxu Ren}, \bibinfo{person}{Beibei Xin}, {and}
  \bibinfo{person}{Yan Liu}.} \bibinfo{year}{2018}\natexlab{}.
\newblock \showarticletitle{Can I trust you more? Model-Agnostic Hierarchical
  Explanations}.
\newblock  (\bibinfo{year}{2018}).
\newblock


\bibitem[\protect\citeauthoryear{Yang, Huang, Scholtz, and Arendt}{Yang
  et~al\mbox{.}}{2020}]%
        {yang2020visual}
\bibfield{author}{\bibinfo{person}{Fumeng Yang}, \bibinfo{person}{Zhuanyi
  Huang}, \bibinfo{person}{Jean Scholtz}, {and} \bibinfo{person}{Dustin~L
  Arendt}.} \bibinfo{year}{2020}\natexlab{}.
\newblock \showarticletitle{How do visual explanations foster end users'
  appropriate trust in machine learning?}. In
  \bibinfo{booktitle}{\emph{Proceedings of the 25th International Conference on
  Intelligent User Interfaces}}. \bibinfo{pages}{189--201}.
\newblock


\bibitem[\protect\citeauthoryear{Yeh, Kim, Yen, and Ravikumar}{Yeh
  et~al\mbox{.}}{2018}]%
        {yeh2018representer}
\bibfield{author}{\bibinfo{person}{Chih-Kuan Yeh}, \bibinfo{person}{Joon Kim},
  \bibinfo{person}{Ian En-Hsu Yen}, {and} \bibinfo{person}{Pradeep~K
  Ravikumar}.} \bibinfo{year}{2018}\natexlab{}.
\newblock \showarticletitle{Representer point selection for explaining deep
  neural networks}.
\newblock \bibinfo{journal}{\emph{Advances in neural information processing
  systems}}  \bibinfo{volume}{31} (\bibinfo{year}{2018}).
\newblock


\bibitem[\protect\citeauthoryear{Yin, Hay, and Roth}{Yin et~al\mbox{.}}{2019}]%
        {yin2019benchmarking}
\bibfield{author}{\bibinfo{person}{Wenpeng Yin}, \bibinfo{person}{Jamaal Hay},
  {and} \bibinfo{person}{Dan Roth}.} \bibinfo{year}{2019}\natexlab{}.
\newblock \showarticletitle{Benchmarking zero-shot text classification:
  Datasets, evaluation and entailment approach}. In
  \bibinfo{booktitle}{\emph{2019 Conference on Empirical Methods in Natural
  Language Processing and 9th International Joint Conference on Natural
  Language Processing, EMNLP-IJCNLP 2019}}. Association for Computational
  Linguistics, \bibinfo{pages}{3914--3923}.
\newblock


\bibitem[\protect\citeauthoryear{Zhou, Gandomi, Chen, and Holzinger}{Zhou
  et~al\mbox{.}}{2021}]%
        {zhou2021evaluating}
\bibfield{author}{\bibinfo{person}{Jianlong Zhou}, \bibinfo{person}{Amir~H
  Gandomi}, \bibinfo{person}{Fang Chen}, {and} \bibinfo{person}{Andreas
  Holzinger}.} \bibinfo{year}{2021}\natexlab{}.
\newblock \showarticletitle{Evaluating the quality of machine learning
  explanations: A survey on methods and metrics}.
\newblock \bibinfo{journal}{\emph{Electronics}} \bibinfo{volume}{10},
  \bibinfo{number}{5} (\bibinfo{year}{2021}), \bibinfo{pages}{593}.
\newblock


\end{thebibliography}
